\documentclass[journal, english, onecolumn, 11pt]{IEEEtran}
\IEEEoverridecommandlockouts

\usepackage[mathscr]{eucal}
\usepackage[utf8]{inputenc} %-- pour utiliser des accents en franÃ¯Â¿Åais, ou autres
\usepackage{%
	algorithm,
	algorithmic,%
	amsmath,%
	amssymb,%
	amsthm,%
	array,%
	booktabs,%
	cite,
	bbm,%
	dsfont,%
	graphicx,%
	multirow,
	paralist,%
	pgf,%
	pgfgantt,%
	pgfplots,%
	siunitx,%
	subcaption,%
	subfiles,%
	tabulary,
	tikz,%
	url,%
	xcolor,%
}
\usepackage[normalem]{ulem}

\usepackage{scalerel,stackengine}

\graphicspath{{Figures/}{../Figures/}}

\usepackage[hidelinks,%
colorlinks	= true,%
urlcolor   	= blue, %Colour for external hyperlinks
linkcolor  	= magenta, %Colour of internal links
citecolor   	= red %Colour of citations
]{hyperref}

\usepgflibrary{shapes}
\usetikzlibrary{%
	arrows,%
	automata,%
	calc,%
	chains,%
	patterns,%
	positioning,%
	shapes,%
	shapes.callouts,%
	shapes.multipart,%
	shapes.symbols,%
}

\renewcommand{\le}{\leqslant}
\renewcommand{\ge}{\geqslant}

\newcommand{\EQ}[1]{\begin{equation*}#1\end{equation*}}
\newcommand{\EQN}[1]{\begin{equation}#1\end{equation}}
\newcommand{\eq}[1]{\begin{align*}#1\end{align*}}
\newcommand{\eqn}[1]{\begin{align}#1\end{align}}

\newcommand{\set}[1]{\left\{#1\right\}}

\newcommand{\SetIn}[1]{\mathbbm{1}_{\set{#1}}}
\newcommand{\E}[1]{\mathbb{E}\left[#1\right]}
\newcommand{\abs}[1]{\left\lvert #1\right\rvert}

\newcommand{\given}{\Big\lvert}
\newcommand{\ceil}[1]{\left\lceil  #1\right\rceil}
\newcommand{\floor}[1]{\left\lfloor  #1\right\rfloor}
\newcommand{\Var}{\operatorname{Var}}
\newcommand{\rep}{\mathrm{rep}}
\newcommand{\mds}{\mathrm{mds}}
\newcommand{\iid}{\emph{i.i.d.~}}

\newcommand{\F}{\mathbb{F}}
\newcommand{\N}{\mathbb{N}}
\newcommand{\R}{\mathbb{R}}
\newcommand{\Z}{\mathbb{Z}}

\newcommand{\cB}{\mathcal{B}}
\newcommand{\cI}{\mathcal{I}}
\newcommand{\cP}{\mathcal{P}}
\newcommand{\cS}{\mathcal{S}}

\newcommand{\fV}{\mathsf{V}}
\newcommand{\bl}[1]{{\color{blue}#1}}
\newcommand{\red}[1]{{\color{red}#1}}

\newcommand{\nix}[1]{}

\nix{
	\usepackage[hidelinks,colorlinks=true,urlcolor=blue, %Colour for external hyperlinks
	linkcolor=blue, %Colour of internal links
	citecolor=red %Colour of citations
	]{hyperref}
}

\theoremstyle{plain}
\newtheorem{thm}{Theorem}
\newtheorem{cor}[thm]{Corollary}
\newtheorem{lem}[thm]{Lemma}
\newtheorem{prop}[thm]{Proposition}

\theoremstyle{definition}
\newtheorem{defn}[thm]{Definition}

\theoremstyle{plain}  
\newtheorem{innercustomthm}{Problem}
\newenvironment{prob}[1]
  {\renewcommand\theinnercustomthm{#1}\innercustomthm}
  {\endinnercustomthm}
  
\theoremstyle{remark}
\newtheorem{rem}{Remark}
\newtheorem{note}{Note}
 
\theoremstyle{plain}

\title{Latency optimal storage and scheduling of replicated fragments for memory-constrained servers}
\author{ 
Rooji~Jinan\IEEEauthorrefmark{1}\quad \quad%\IEEEmembership{Student~Member,~IEEE},\quad %
\and Ajay~Badita\IEEEauthorrefmark{2}\quad \quad~%\IEEEmembership{Student~Member,~IEEE},\quad %
\and Pradeep~Sarvepalli\IEEEauthorrefmark{3}\quad \quad~%\IEEEmembership{Member,~IEEE}, \quad%
\and Parimal~Parag\IEEEauthorrefmark{2}~%\IEEEmembership{Member,~IEEE}%
\thanks{
Author\IEEEauthorrefmark{1} is with the Robert Bosch centre for cyber-physical systems, 
and the authors\IEEEauthorrefmark{2} are with the department of electrical communication engineering, 
all at Indian Institute of Science, Bangalore, Karnataka 560012, India. 
Email:\IEEEauthorrefmark{1}\{roojijinan, ajaybadita, parimal\}@iisc.ac.in.
}
\thanks{
Author\IEEEauthorrefmark{2} is with the department of electrical engineering, 
Indian Institute of Technology Madras, Tamil Nadu 600036, India. 
Email:\IEEEauthorrefmark{2}\{pradeep\}@ee.iitm.ac.in. 
}
}

\begin{document}
	\maketitle
	\begin{abstract}
We consider the setting of distributed storage system where a single file is subdivided into smaller fragments of same size which are then replicated with a common replication factor across servers of identical cache size. 
An incoming file download request is sent to all the servers, 
and the download is completed whenever request gathers all the fragments. 
At each server, we are interested in determining the set of fragments to be stored, 
and the sequence in which fragments should be accessed,  
such that the mean file download time for a request is minimized. 
We model the fragment download time as an exponential random variable independent and identically distributed for all fragments across all servers, 
and show that the mean file download time can be lower bounded in terms of the expected number of useful servers summed over all distinct fragment downloads. 
We present deterministic storage schemes that attempt to maximize the number of useful servers. 
We show that finding the optimal sequence of accessing the fragments is a Markov decision problem,
whose complexity grows exponentially with the number of fragments. 
We propose heuristic algorithms that determine the sequence of access to the fragments which are empirically shown to perform well. 
\end{abstract}

	%!TeX spellcheck = en_US
%!TeX spellcheck = LaTeX

\section{Introduction}
The recent years have seen a widespread deployment of distributed storage systems, consisting of large number of storage nodes. 
These nodes are prone to failures and unpredictable download times~\cite{Dimakis2010TIT}. 
A primary challenge in such systems is to provide resilience against such events.  
One approach for improving the robustness of a distributed storage system is  adding redundancy through error-correcting codes. 
It turns out that these systems also offer fast access to the data due to parallelization gains. 
Latency redundancy trade-off has been studied for maximum distance separable (MDS) and replication codes in various articles~\cite{Shah2016TCOM, Li2016Infocom, Lee2017TNET, Gardner2017TNET, Parag2017Infocom, Lee2018TIT}. 

%\begin{defn} 
For simplicity, we consider a single file of unit size. 
A distributed storage system of $B$ servers that can each store $\alpha$ fraction of this file, 
is referred to as an $\alpha$-$B$ system. 
A single file of unit size divided into $V$ fragments,  
encoded into $VR$ fragments, 
and stored over an $\alpha$-$B$ system,
is called an $\alpha$-$(B, V, R)$ coded storage scheme.  
%\end{defn} 
This suggests that we can use $(n,k)$ error-correcting codes which encode $k=V$ information symbols into 
$n=VR$ encoded symbols, 
such that the file can be decoded by downloading certain coded symbols. 
In replication coding, 
we replicate each of the $V$ file fragments $R$ times, 
and the file can be decoded by downloading a single replica of each fragment. 
On the other hand, in a $(VR,V)$ MDS code the $V$ file fragments are encoded into $VR$ fragments, 
and the entire file can be reconstructed from any of the $V$ encoded fragments. 

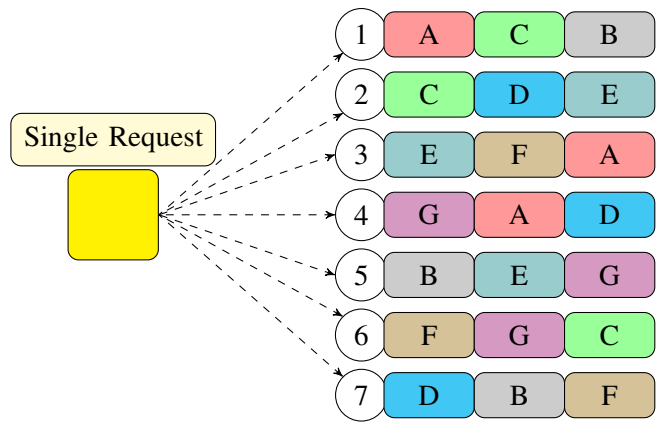
\begin{figure}[hhh]
\centerline{\scalebox{1}{\begin{tikzpicture}
[node distance=2cm, draw=black, thin, >=stealth', 
  req/.style={rectangle, rounded corners, minimum size=12mm, draw=black},
  vault/.style={circle, minimum size = 2cm, draw=black},
  server/.style={circle, minimum size = 0.5cm, draw=black},
  content/.style={rectangle, rounded corners, minimum height=1cm, minimum width=4cm, draw=black},
  cache/.style={rectangle, rounded corners, minimum height=0.6cm, minimum width=1.2cm, draw=black, inner sep=2pt},
  note/.style={rectangle, rounded corners, minimum size=5mm, inner sep=5pt, draw=black}
]	
%----------------------First Stage---------------------------
  \node[server](server1) at (0.1,4.8){1};
  \node[cache, fill=red!40] (cache11) at (1, 4.8) {A};
  \node[cache, fill=green!40] (cache12) at (2.2, 4.8) {C};
  \node[cache, fill=black!20] (cache13) at (3.4, 4.8) {B};

  \node[server](server2) at (0.1,4){2};
  \node[cache, fill=green!40] (cache21) at (1, 4) {C};
  \node[cache, fill=cyan!60] (cache22) at (2.2, 4) {D};
  \node[cache, fill=blue!50!green!40] (cache23) at (3.4, 4) {E};

  \node[server](server3) at (0.1,3.2){3};
  \node[cache, fill=blue!50!green!40] (cache31) at (1, 3.2) {E};
  \node[cache, fill=green!40!red!40] (cache32) at (2.2, 3.2) {F};
  \node[cache, fill=red!40] (cache33) at (3.4, 3.2) {A};

  \node[server](server4) at (0.1,2.4){4};
  \node[cache, fill=blue!40!red!40] (cache41) at (1, 2.4) {G};
  \node[cache, fill=red!40] (cache42) at (2.2, 2.4) {A};
  \node[cache, fill=cyan!60] (cache43) at (3.4, 2.4) {D};

  \node[server](server5) at (0.1,1.6){5};
  \node[cache, fill=black!20] (cache51) at (1, 1.6) {B};
  \node[cache, fill=blue!50!green!40] (cache52) at (2.2, 1.6) {E};
  \node[cache, fill=blue!40!red!40] (cache53) at (3.4,1.6) {G};
  
  \node[server](server6) at (0.1,0.8){6};
  \node[cache, fill=green!40!red!40] (cache61) at (1, 0.8) {F};
  \node[cache, fill=blue!40!red!40] (cache62) at (2.2, 0.8) {G};
  \node[cache, fill=green!40] (cache63) at (3.4, 0.8) {C};
  
  \node[server](server7) at (0.1,0){7};
  \node[cache, fill=cyan!60] (cache71) at (1, 0) {D};
  \node[cache, fill=black!20] (cache72) at (2.2, 0) {B};
  \node[cache, fill=green!40!red!40] (cache73) at (3.4, 0) {F};

  \node[req,fill=yellow] (req01) at (-3.2,2.4){};
  \node[note, fill=yellow!20] at (-3.2,3.4) {Single Request};
  
  \path (server1.south west) edge[<-, dashed] (req01.east);
  \path (server2.south west) edge[<-, dashed] (req01.east);
  \path (server3.west) edge[<-, dashed] (req01.east);
  \path (server4.west) edge[<-, dashed] (req01.east);
  \path (server5.west) edge[<-, dashed] (req01.east);
  \path (server6.north west) edge[<-, dashed] (req01.east);
  \path (server7.north west) edge[<-, dashed] (req01.east);

\end{tikzpicture}}}
\caption{
An example of $\frac{3}{7}$-$(7,7,3)$ replication coded system, 
where a single file is fragmented into $V=7$  fragments, 
and each fragment is repeated $R=3$ times over $B = 7$ servers each with storage of $K=3$ fragments. 
The corresponding block replication code is denoted by $(21, 7)$. 
}
\label{fig:BlockRep}
\end{figure}

For an $\alpha$-$B$ system, 
it is assumed that per server storage is $K = \alpha V$ in terms of the fragments.
From the memory constraint on each of $B$ servers, 
we have $VR \le BK$ or $R \le \alpha B$. 
An $\alpha$-$(B,V,R)$ coded storage scheme is called \emph{completely utilizing} if the encoded $VR$ fragments completely utilize the available memory of the  underlying $\alpha$-$B$ system, 
and \emph{underutilizing} otherwise. 
For completely utilizing and underutilizing $\alpha$-$(B,V,R)$ coded storage schemes, 
we have  $R = \alpha B$ and $R < \alpha B$ respectively.  
Fig.~\ref{fig:BlockRep} shows such a scheme for $\frac{3}{7}$-$(7,7,3)$ replication storage scheme. 

A commonly studied $\alpha$-$(B, V, R)$ system is where per server storage $\alpha=\frac{1}{V}$~\cite{Wu2007Allerton,Dimakis2010TIT,Badita2019TIT}, 
that is the number of coded fragments on each of the $B$ parallel server caches is $K=1$. 
For this case it was shown that MDS codes provide optimal performance~\cite{Badita2019TIT} in terms of the mean access delay. 
However, if we allow for an additional degree of freedom, namely a larger subpacketization of the fragments stored on the servers, 
then even non-MDS codes can become competitive. 
For instance, when $K>1$ the staircase codes proposed in~\cite{Bitar2018TIT} for secure computation
have been used to improve upon the MDS codes. 
The mean download time of staircase codes was shown to be smaller than that of a $(B, B/R)$ MDS code in~\cite{Bitar2017ISIT, Bitar2020TCOM}.
  
Fixing the subpacketization at $V$ fragments, the code rate at $1/R$, 
and the number of servers at $B$, 
it can be shown that among all $(VR, V)$ codes stored over an $\alpha$-$B$ system, 
an MDS code has the smallest file download time for a class of  distributions. 
Even though MDS codes are latency optimal, they have certain drawbacks. 
First, encoding and decoding require complex finite field arithmetic. 
The best known MDS decoding algorithms are polynomial in the number of coded symbols $VR$~\cite{Lin2001ECC}. 
For a slow processing system, this could lead to a non-trivial decoding latency. 
Second, file sizes often change in storage systems with frequent writes~\cite{RenVLDB2013}. 
In this case, the entire file has to be encoded again~\cite{Maturana2020ISIT}. 
Third, to be able to code $V$ fragments of a file into $VR$ MDS coded symbols, 
the symbols must belong to a sufficiently large alphabet~\cite[Theorem 4.1] {Roth2006CodingTheory}.
This can be achieved by grouping multiple bits together in each file fragment. 
This puts a constraint that each fragment should be sufficiently large. 

Block replication codes score well on all of these fronts. 
As the replication codes are binary,  encoding and decoding of replication is trivial and the file sizes can be small.
Further, the file size changes can be accommodated by change in the associated fragments and their replicas. 
This is reflected in widespread adoption~\cite{Pawar2011ISIT, Zhu2014NCC, Wang2015Sigmetrics,Badita2020TNET} of replication codes in distributed systems. 
Furthermore, we will show that the latency performance of block replication codes becomes comparable to that of MDS codes with increase in either  the number of fragments or the storage size per server. 
Hence, we focus on $(VR, V)$ replication codes that offer good mean download time for an $\alpha$-$B$ system. 
For replication codes, we need to determine on which servers each of the replicated fragments should be stored.
In addition, we need to consider the order in which the fragments are accessed at each server. 
{\em We consider the problem of optimal storage and access sequence of the replicated fragments at each of the $B$ servers, 
such that the mean download time is minimized.}
There is no obvious relation between the fragments stored on each of the servers and the mean download time. 
Therefore, determining what fragments should be stored on each of the servers appears to be a difficult problem as there are exponentially many ways to store the fragments on the servers. 
For a given fragment storage on each server, the problem of optimal access sequence can be posed as a finite horizon Markov Decision Problem(MDP),  
which can be solved by standard backward induction algorithm~\cite{Puterman1994Wiley}. 
For the proposed MDP, this algorithm requires exponentially large memory in number of fragments, 
and cannot be implemented efficiently for large system parameters. 

\subsection{Related work}
Coding techniques have emerged as a popular technique to provide reliability in distributed storage systems with fault-prone network~\cite{Dimakis2010TIT, Suh2010ISIT,Rouayheb2010Allerton, Rashmi2013Usenix, Rawat2016TIT}. 
Storage codes can also be designed to achieve additional objectives such as low repair bandwidth~\cite{Wu2007Allerton,Dimakis2010TIT,Shah2010ITW,Pawar2011ISIT,Papailiopoulos2013TIT}, 
low regeneration bandwidth~\cite{el2010Allerton, Rashmi2011TIT,Shum2013TIT},  high locality~\cite{Wang2014TIT,Wang2015ISIT}, 
low latency~\cite{Dean2013CACM,Joshi2014JSAC,Liang2014Allerton,Joshi2015Allerton }, among others. 
In this work, we are interested in low latency performance of distributed storage systems, 
by using codes.
Specifically, we study replication codes 
where files are stored redundantly over the system. 

It has been shown that redundant storage can reduce latency as well. 
In this case, a download request can be served in parallel by multiple servers storing the requested file~\cite{Badita2019TIT,Vulimiri2013ACM,Joshi2015Allerton,Joshi2015Sigmetrics,Gardner2015ACM,Xiang2016TNET,Shah2016TCOM,Parag2017Infocom,Parag2018ITA}. 
Trade-off between latency and cost of availing redundancy was observed empirically in~\cite{Vulimiri2013ACM}, 
and subsequently studied theoretically in~\cite{Joshi2014JSAC,Joshi2015Sigmetrics,Xiang2016TNET,Parag2017Infocom,Badita2020INFOCOM,Badita2020TNET}. 
Two well studied file encoding strategies used in distributed systems with redundant storage are MDS coding~\cite{Suh2010ISIT,Joshi2014JSAC,Badita2019TIT, Badita2020INFOCOM} and  replication~\cite{Rouayheb2010Allerton, Zhu2014NCC, Wang2015Sigmetrics, Badita2020TNET}. 
It has been shown that MDS coding outperforms replication in mean file access latency~\cite{Huang2012ISIT, Li2016Infocom, Badita2019TIT}. 

In all these works, it was assumed that each server stores a single coded fragment of the file. 
In other words, the size of a file fragment is equal to the memory of the server. 
If we divide the file into smaller fragments than the memory available at the server, 
then we can store multiple file fragments on each server. 
This fact was exploited by~\cite{Bitar2017ISIT, Bitar2020TCOM} to show that the mean access latency for Staircase codes can be smaller than MDS codes storing single coded fragment stored on each server. 
In Staircase codes, the fragments stored on the servers are not all of equal size. 
We note that if the file is uniformly subfragmented and a larger code is used, 
then performance of MDS codes can also be improved.

Our work differs from these existing works in the following ways. 
We focus on replication codes for equal sized fragments of a single file stored over multiple servers, 
such that each server can store multiple fragments. 
Each server stores a different set of fragments introducing an asymmetry among the servers.
The storage scheme has a direct impact on the mean download time. 
Furthermore, the sequence of access of fragments at each server affects the download time significantly.
We address the problem of constructing good storage scheme  as well as the access sequence at each server in this work.
These aspects have not been explored in the literature to the best of our knowledge. 

\subsection{Main contributions}
We briefly summarize the main contributions of the paper.

\begin{compactenum}[i)]
\item We study replication codes for file fragments stored over  distributed storage systems.
Our study differs from previous work in that each server can store multiple fragments. 

\item We characterize the mean download time of a file  
when fragment download times are random and independent and identically distributed(\emph{i.i.d}) exponential, 
and find a lower bound in terms of the expected sum of number of useful servers for each fragment download. 

\item We provide bounds on the number of useful servers for any fragment storage scheme in an $\alpha$-$B$ system employing $(VR,V)$ replication code.  

\item We propose fragment storage schemes that maximize the aforementioned lower bounds on number of useful servers. 

\item We establish that finding the optimal fragment access sequence is an MDP. 
We propose efficient suboptimal algorithms that are easy to implement.

\item We show that among all the $(VR,V)$ codes stored on an $\alpha$-$B$ system, 
an MDS code minimizes the mean download time. 
In addition, we show that a $(VR,V)$ replication code matches the MDS code performance when $\alpha \ge 1$, i.e. $K \ge V$.  

\item We propose a random fragment storage scheme for $(VR,V)$ replication codes
that performs competitively with respect to $(VR,V)$ MDS codes when $K<V$, for large $V$.

\item We support our analyses with numerical studies which provide additional insights into $\alpha$-$(B, V, R)$ coded storage schemes.
They also illustrate the performance of various proposed algorithms.  

\end{compactenum}

A key takeaway from our work is that a good choice of storage scheme and access sequence can enable replication codes to be a practical alternative to MDS codes in certain situations.
For instance, 
when files are updated frequently, 
or when encoding/decoding complexity is comparatively expensive, 
or when file size is small. 

\subsection{Organization}
We present the system model in Section~\ref{sec:model} and the problem formulation in Section~\ref{sec:ProblemForm}. 
We provide universal performance bounds in Section~\ref{sec:bounds}, 
based on which we propose deterministic storage policies in Section~\ref{sec:DesignBasedStorage}. 
Algorithms for fragment access sequence are studied in Section~\ref{sec:Scheduling}. 
A random storage scheme for replication codes is presented in Section~\ref{sec:RandomCode}. 
The performance of replication and MDS codes are compared in Section~\ref{sec:ComparisonMDS}. 
Empirical studies are provided in Section~\ref{sec:sims}, 
and we conclude with some final remarks in Section~\ref{sec:Conclusion}.

{\em Notation.}
We briefly summarize the notation used throughout this article.  
We denote the set of first $N$ consecutive positive integers by $[N] \triangleq \set{1, \dots, N}$, 
the set of positive integers by $\N$, 
the set of non-negative integers by $\Z_+$, 
the set of non-negative reals by $\R_+$. 
For a set $A$, 
we denote the collection of all subsets by $2^A$,
and the cardinality by $\abs{A}$. 

	% !TEX spellcheck = en_US
% !TEX spellcheck = LaTeX

\section{System model} 
\label{sec:model}
We consider the storage of a single file fragmented into $V$ pieces on a finite number of servers $B$. 
Each of these servers are assumed to have identical storage capacity of $K$ such fragments.  
We will initially consider $K \le V$, and define the storage capacity per server in terms of fraction of file fragments 
\EQN{
\alpha \triangleq \frac{K}{V}. 
}
We will see in Section~\ref{subsec:LargeStorage}, that $K\ge V$ is a special case and can be studied independently. 
For $(VR,V)$ block codes that encode $V$ fragments into $VR$ coded fragments, 
the \emph{code rate} is $1/R$. 
We study $(VR, V)$ replication codes  
where each file fragment is replicated $R$ times, 
and $R$ is called the {\em replication factor} of the code. 
The system should have sufficient storage capacity to store all $VR$ fragments, 
and this requires that $VR\le KB$, 
or equivalently $R\le B\alpha$. 

We assume a single request in the system which is forked to all $B$ servers. 
At each server $b \in [B]$,  the request starts downloading stored fragments in succession. 
The fragment to be downloaded at each server depends on the fragments that have already been downloaded. 
Any server that contains a fragment that has not yet been downloaded is called a \emph{useful server}. 
The download time for the file is affected by which  fragments are stored on each server, 
download time of each fragment,  
and the temporal sequence in which the fragments are downloaded. 
In the following sections, we consider each of these aspects in more detail. 

\subsection{Storage model}
In the context of distributed storage systems it is not enough to specify the storage code,
one also needs to specify where the coded fragments are stored. 
For a specific storage code, we refer to the collection of fragments stored on each server as the \emph{storage scheme}. 
The storage scheme must be designed with a view to facilitate fast download of the file. 

For replication codes, some storage schemes are suboptimal.
For instance, when  multiple copies of a fragment are stored on the same server, there
is no parallelization gain. 
This scheme implicitly under-utilizes the storage on each server. 
Therefore, we study storage schemes where each server stores at most a single copy of each fragment. 
\begin{defn}[Occupancy set]
\label{defn:OccupancySet}
The set of servers on which a fragment $v \in [V]$ is replicated, is denoted by $\Phi_v \subseteq [B]$ and called \emph{occupancy set}.  
\end{defn}
Recall that each fragment is replicated an identical number of times in the system, 
and the replication factor $R = \abs{\Phi_v}$ for all fragments $v \in [V]$.  
The storage scheme is completely determined by the collection of occupancy sets $\Phi \triangleq (\Phi_v: v \in [V])$. 

\begin{defn}[Fragment set]
\label{defn:FragmentSet}
We denote the set of fragments stored on a server $b \in [B]$ by the \emph{fragment set} $S_b \subseteq [V]$, 
such that 
\EQN{
\label{eqn:FragmentSet}
S_b \triangleq \set{v \in [V]: b \in \Phi_v}.
}
\end{defn}

Recall that we classify the storage of $(VR,V)$ codes on $\alpha$-$B$ systems as \emph{completely utilizing} and \emph{underutilizing}. 
In a \emph{completely utilizing} storage scheme, 
the storage capacity at each server is completely used, i.e. $\abs{S_b} = K$ for all servers $b \in [B]$. 
In such schemes $VR = BK$, 
since there are $R$ replications of $V$ fragments on $B$ servers that can store $K$ fragments each. 
This also implies that the number of servers $B$ can be written in terms of the fraction $\alpha$ of file fragments stored per server and the replication factor $R$ as $B = \frac{R}{\alpha}$.
A $(VR,V)$ replication code stored on a $\alpha$-$B$ system is said to be \emph{underutilizing}, if it is not completely utilizing.
For such schemes, the replication factor $R < \alpha B$.

\begin{defn}
\label{defn:BVRKStorage} 
For any $(VR,V)$ replication code stored on $B$ servers with storage capacity of $K$ fragments, 
the \emph{completely utilizing $\alpha$-$(V,R)$ replication storage ensemble} is defined as the collection 

\EQN{
\label{eqn:BVRKStorage} 
\cS \triangleq \set{\Phi \in (2^{[B]})^{[V]}\Bigg| \begin{array}{l}{\abs{\Phi_v} = R, \abs{S_b} = K }\\
{\text{for all }v, b, VR = BK}\end{array} }.
}

\end{defn}
Let us also elaborate on the special case of underutilizing $\alpha$-$(V,R)$ replication storage scheme.
An underutilizing $\alpha$-$(V,R)$ replication storage scheme satisfies $\abs{S_b} = K$ but allows multiple copies of the same fragment to be stored on the same server.
That is, $\abs{\Phi_v} < R = \alpha B$.
In short, it completely utilizes the storage available at each server, but it is still underutilizing in the sense that it do not take full advantage of the possible parallelization.

Once a storage scheme $\Phi$ from $\cS$ has been chosen, the fragments are placed on the servers according to this scheme. 
File download time in an $\alpha$-$(V,R)$ system is affected not only by the storage scheme $\Phi$ in $\cS$, but also on the fragment download times,   
and the sequence in which the fragments are downloaded. 
We consider these next. 

\subsection{Fragment download time model}

The fragment download time at each server  
is modeled by a random variable that captures uncertainty due to network delays and server background processes~\cite{Cheng2014IMC}.   
We denote the fragment download time for fragment $v$ at the server $b$ by a nonnegative random variable $T_{bv}$. 
We assume that all fragments are of equal size and the servers are identical,  
in the sense that the marginal distribution of $T_{bv}$ is identical for all fragments $v \in S_b$ at all servers $b \in [B]$. 
Motivated by analytical tractability, we further assume that the fragment download times $T_{bv}$ are independent. 
Thus, we have assumed fragment download times to be \emph{i.i.d.}, 
which is a popular assumption in the literature~\cite{Xiang2016TNET,Shah2016TCOM,Wang2015Sigmetrics,al2019ttloc,Lee2017TNET,Badita2019TIT}. 

It has been shown that shifted exponential distribution is a good model for the random download time in data center networks~\cite{Lee2018TIT, Bitar2017ISIT, Xiang2016TNET, Al2018TNET}, 
where the constant shift is the startup time for servers, and the memoryless part accounts for the uncertainty. 
In the case when the startup time is negligible when compared to the mean download time, 
exponential distribution is a good approximation for the download time distribution. 
Therefore, we also assume the common distribution function $F$ for the random fragment download time to be exponentially distributed with rate $\mu$, 
such that
\EQ{
F(x) \triangleq P(\set{T_{bv} \le x})= 1-e^{-\mu x},~\text{ for all } x \ge 0.
} 

\subsection{Download sequence and scheduling}

Recall that the request is completely serviced when all the $V$ distinct fragments have been downloaded. 
Due to the stochastic nature of fragment download, 
the order in which the request downloads the fragments is random. 
We assume that the request only downloads the fragments which have not been downloaded before, 
and such download sequences are referred to as \emph{minimal downloading sequences}. 
A minimal downloading sequence for replication codes consists of unique fragments. 
The $\ell$th downloaded fragment is denoted by $v_{\ell}$, 
and the sequence of downloaded fragments until $\ell$th download is called download subsequence and denoted by $I_\ell \triangleq (v_1, \dots, v_\ell)$. 
Since all the downloaded fragments are unique, 
sometimes we regard the download subsequence $I_\ell$ as a set and ignore the ordering of fragments. 
The distinction between the set and the subsequence will be clear from the context. 

Given the sequence of downloaded fragments, the process of 
determining which fragment is made available for downloading at each server is referred to as 
{\em scheduling}. 
We will only be interested in class of scheduling policies that result in minimal download sequences, 
and refer to them as \emph{work-conserving}. 
These scheduling policies do not waste the servers' work on downloading replicas of the fragments that have already been downloaded. 
In other words, the scheduler only selects the the remaining fragments at each server.  
Therefore, 
after a fragment download, 
the request immediately stops downloading that fragment from other servers. 
Instead, it starts downloading the scheduled fragment among one of the remaining fragments at that server. 
After $\ell$th download, the set of remaining fragments on a server $b \in [B]$ is denoted as 
\EQN{
\label{eqn:ResidualFragmentSet}
S_b^{\ell}\triangleq S_b\setminus I_\ell,\quad \ell \in [V].
}

\begin{defn}[Work-conserving scheduling]
\label{defn:Scheduling}
The work-conserving scheduling can be formally defined as a function $\Psi: \cS \times 2^{[V]} \to [V]^{[B]}$, 
that selects one of the remaining fragments at each server $b$ after $\ell$ downloads for a storage scheme $\Phi \in \cS$. 
That is, for all $\ell \in \set{0, \dots, V-1}$, 
we have 
\EQN{
\label{eqn:DefinitionScheduling}
\Psi(\Phi, I_\ell)(b) \in S_b^\ell,~\text{ for all } b \in 
[B] \text{ such that } S_b^\ell\neq \emptyset.
}
We denote the restriction of work-conserving scheduling policy $\Psi$ to a fixed storage scheme $\Phi \in \cS$ by $\Psi_\Phi: 2^{[V]} \to [V]^{[B]}$ such that 
\EQ{
\Psi_\Phi(I_\ell)(b) = \Psi(\Phi,I_\ell)(b).
}
\end{defn}

Motivated by analytical tractability, we have assumed an idealized model of perfect and immediate cancellation at all parallel servers serving the same fragment. 
We further remark that cancellation delays would impact any redundant access scheme, where remaining redundant requests have to be cancelled. 
This has been studied in~\cite{Lee2017TNET}. 
Since our main objective is to show that replication codes can be competitive to other storage codes, 
we ignore this cancellation delay. 

	\section{Problem formulation}
\label{sec:ProblemForm}

In this section, we give a precise formulation of the main problem we study.
Our main goal is to minimize the mean download time for a file stored using a storage scheme $\Phi$ from the completely utilizing $\alpha$-$(V,R)$
replication storage ensemble and work-conserving scheduling scheme $\Psi$, 
when the service times at each server are independent and exponentially distributed with rate $\mu$. 
Recall that each file is divided into $V$ fragments prior to storage, 
and a unique replica of each fragment has to be downloaded for a successful file download. 

We denote the download time of $\ell$th distinct fragment $v_\ell$ by $D_\ell$, where $D_0 \triangleq 0$ and $\ell \in \set{0, \dots, V}$. 
This indicates that a download subsequence $I_\ell$ of $\ell$ fragments has been downloaded at time $D_\ell$. 
With this notation, the file download time is denoted by $D_V$. 
\nix{The random sequence $(I_0, \dots, I_V)$ of fragment downloads is called the \emph{information sequence}. 
There is no new fragment download during the $(\ell+1)$th download interval $[D_\ell, D_{\ell+1})$, 
and the information subsequence remains $(I_0, \dots, I_\ell)$ in this time duration. }

If the request has downloaded all the fragments available at a server, the corresponding sever is called \emph{useless} server. 
Any server that has fragments not yet downloaded by the request, is called a \emph{useful server}\footnote{We borrow this nomenclature of useful and useless servers from~\cite{Badita2019TIT}.}. 
The set of useful servers after $\ell$th download is denoted $U(I_\ell)$ and its cardinality by 
$N(I_\ell)$. 
More precisely, we have  
\begin{xalignat}{2}
\label{eqn:UsefulServers}
&U(I_\ell) = \bigcup_{v \notin I_\ell}\Phi_v,&
&N(I_\ell) = \abs{U(I_\ell)}. 
\end{xalignat}

The request is being served by $N(I_\ell)$ parallel servers in the duration $[D_\ell, D_{\ell+1})$.  
From the independent and memoryless service assumption at all servers of rate $\mu$, 
we have 
\EQ{
\E{D_{\ell+1}-D_\ell\given I_\ell} =  \frac{1}{N(I_\ell)\mu}. 
}
The download time for $V$ fragments can be written as the sum of download time of individual fragments, i.e. $D_V = \sum_{\ell=0}^{V-1}(D_{\ell+1}-D_\ell)$. 
From the linearity and the tower property of expectation, 
it follows that the mean download time averaged over all fragments is 
\EQN{
\label{eq:MeanDownloadTime}
\frac{1}{V}\E{D_V} 
= \frac{1}{V}\E{\sum_{\ell=0}^{V-1}\frac{1}{N(I_\ell)\mu}}.
}

We see that the mean download time depends on 
$U(I_\ell)$, the set of useful servers  remaining after $\ell$th download, 
which in turn depends on the storage scheme $\Phi$ and the scheduling policy $\Psi$. 
We present the following lemma that provides us a lower bound on the mean file download time in terms of the sum of mean number of useful servers.  
\begin{lem}
For any positive random vector $X \in \R_+^V$, we have 
\EQ{
\frac{1}{V}\E{\sum_{i=1}^V\frac{1}{X_i}} \ge \frac{V}{\sum_{i=1}^{V}\E{X_i}}.
}
\end{lem}
\begin{IEEEproof}
Recall that the arithmetic mean is always larger than the harmonic mean, and hence we can write $\frac{1}{V}\sum_{i=1}^V\frac{1}{X_i} \ge V/(\sum_{i=1}^VX_i)$. 
Taking expectation on both sides, and applying Jensen's inequality~\cite{Coverbook2006} to the convex function $f(x) = \frac{1}{x}$ and positive random variable $\sum_{i=1}^{V}X_i$, we get the result. 
\end{IEEEproof} 
\begin{rem}
It follows from the above lemma that the mean download time can be lower bounded as 
\EQN{
\label{eqn:MeanDownloadTimeLowerBound}
\frac{1}{V}\E{D_V} \ge \frac{V}{\mu\sum_{\ell=0}^{V-1}\E{N(I_\ell)}}.
}
\end{rem} 
In some important settings, the sum $\sum_{\ell}\E{N(I_\ell)}$ is analytically more tractable when compared to $\sum_{\ell=0}^{V-1}\E{{1}/{N(I_\ell)}}$. 
Motivated by this fact, instead of minimizing the mean download time in Eq.~\eqref{eq:MeanDownloadTime}, 
we minimize the lower bound in Eq.~\eqref{eqn:MeanDownloadTimeLowerBound}, which is then equivalent to maximizing  $\sum_{\ell=0}^{V-1}\E{N(I_\ell)}$. 
Guided by this observation, we now pose the following problem. 
\begin{prob}{1}
\label{prob:MeanServiceCompletionTime} 
Find the storage scheme $\Phi$ in completely utilizing $\alpha$-$(V,R)$ replication storage ensemble 
and restriction $\Psi_\Phi$ of work-conserving scheduling policy $\Psi$ to this storage scheme, 
that maximizes the mean number of useful servers averaged over all fragments, i.e. 
\EQ{
(\Phi^{\ast}, \Psi^\ast_{\Phi^\ast}) = \arg\max_{(\Phi, \Psi_\Phi)}\frac{1}{V}\sum_{\ell=0}^{V-1}\E{N(I_\ell)}.
}
\end{prob}
That is, we must find the optimal storage scheme along with the 
restricted scheduling policy that maximizes the mean number of useful servers averaged over all fragments.  
We divide this problem into two subproblems. 
The first subproblem is to find the optimal scheduling policy given a fixed storage scheme.
\begin{prob}{1-A} 
\label{prob:OptSched}
Find the optimal work-conserving scheduling policy restricted to a fixed completely utilizing $\alpha$-$(V,R)$ replication storage scheme $\Phi$, i.e.
\EQ{
\Psi^\ast_\Phi = \arg\max_{\Psi_\Phi} \frac{1}{V}\sum_{\ell=0}^{V-1}\E{N(I_\ell)}.
}
The optimal work-conserving scheduling policy $\Psi^\ast$ is the collection of restrictions $\Psi^\ast_\Phi$ for all storage schemes $\Phi \in \cS$. 
\end{prob}

The second subproblem is to find the optimal storage scheme given a fixed scheduling policy.
\begin{prob}{1-B} 
\label{prob:OptPlace}
Given a work-conserving scheduling policy $\Psi$,  
find the optimal completely utilizing $\alpha$-$(V,R)$ replication storage scheme $\Phi \in \cS$, i.e.
\EQ{
\Phi^\dagger(\Psi) = \arg\max_{\Phi \in \cS} \frac{1}{V}\sum_{\ell=0}^{V-1}\E{N(I_\ell)}.
}
\end{prob}
By solving Problem~\ref{prob:OptSched} for each storage scheme $\Phi \in \cS$, 
we can find the optimal work-conserving scheduling policy $\Psi^\ast$.   
Subsequently, 
we can obtain the optimal storage scheme  $\Phi^\ast = \Phi^\dagger(\Psi^\ast)$ 
by solving Problem~\ref{prob:OptPlace} for the optimal scheduling $\Psi^\ast$.  
It follows that if we can solve the above two sub-problems, 
then we can find the optimal solution $(\Phi^\ast, \Psi^\ast_{\Phi^\ast})$ to Problem~\ref{prob:MeanServiceCompletionTime}. 

It turns out that Problem~\ref{prob:OptSched} can be posed as an MDP, for a given storage scheme $\Phi$. 
This MDP suffers from the well-known curse of dimensionality~\cite{Bellman1957Princeton},
and becomes intractable for large values of number of fragments. 
We propose heuristic scheduling algorithms that are computationally efficient. 
These algorithms are empirically shown to have a good performance for given storage schemes.  

In contrast to Problem~\ref{prob:OptSched}, it is not clear at the outset, 
how to efficiently solve Problem~\ref{prob:OptPlace}. 
A brute force way of solving this problem would be to compute the mean download time for all storage schemes $\Phi \in  \cS$ under the optimal scheduling policy $ \Psi^\ast_\Phi$, or its surrogate suboptimal heuristic algorithms, 
and searching among all storage schemes. 
Since this brute force search is computationally expensive, 
we propose an alternative suboptimal approach. 
This suboptimal selection of storage scheme together with heuristic scheduling algorithms, is empirically shown to have good performance. 

Instead of directly maximizing the mean number of useful servers, we first find universal lower bounds on the number of useful servers which are agnostic to scheduling algorithm and then maximize these lower bounds.
These lower bounds depend only on certain properties of storage schemes $\Phi \in \cS$. 
Good storage schemes can be chosen to maximize the universal lower bounds on the aggregate number of useful servers. 
This suboptimal selection of storage scheme together with heuristic scheduling algorithms, is empirically shown to have good performance. 

	\section{Performance bounds} \label{sec:bounds}
In this section, 
we study the aggregate number of useful servers for completely utilizing $\alpha$-$(V,R)$ replication storage schemes $\Phi \in \cS$, 
in conjunction with a work-conserving scheduling policy $\Psi$. 
In particular, 
we provide bounds on the number of useful servers $N(I_\ell)$ after each download $\ell \in \set{0, \dots, V-1}$. 
We observe that these bounds are independent of the work-conserving scheduling policies $\Psi$.
The upper bound holds universally for any completely utilizing $\alpha$-$(V,R)$ replication storage scheme $\Phi \in \cS$. 
The lower bound depends on certain properties of the storage scheme $\Phi$. 
Recall that our goal is to maximize the mean of aggregate number of useful servers. 
Motivated by this fact, we find storage schemes that maximize the lower bound on the aggregate number of useful servers.

\subsection{Upper bound on $N(I_\ell)$} 
First, we prove a simple upper bound for the number of useful servers for any $\alpha$-$(V,R)$ storage scheme.

\begin{thm}
\label{thm:upperBoundGen}
For a completely utilizing $\alpha$-$(V,R)$ replication storage scheme $\Phi \in \cS$ defined in Eq.~\eqref{eqn:BVRKStorage}, the number of useful servers $N(I_\ell)$ after $\ell$ downloads is upper bounded in terms of $m \triangleq \ceil{B/R}$, as
\EQN{
\label{eq:UB}
N(I_\ell) \le B\SetIn{\ell \le V-m} + (V-\ell)R\SetIn{\ell > V-m}.
}
\end{thm} 
\begin{IEEEproof} 
From the construction of a completely utilizing $\alpha$-$(V,R)$ storage scheme, 
it follows that all servers are useful before any download is initiated, i.e. $N(I_0) = B$. 
Further, no servers are useful after all fragments have been downloaded, i.e. $N(I_V) = 0$. 
Taking cardinality of the set of useful servers given in Eq.~\eqref{eqn:UsefulServers}, 
we get 
\EQ{
N(I_\ell) 
= \abs{\cup_{v \notin I_\ell}\Phi_v} 
\le \sum_{v \notin I_\ell}\abs{\Phi_v} 
=(V - \ell)R. 
}
Since the number of useful servers cannot exceed the total number of servers $B$, we get 
$N(I_\ell) \le \min\set{B, (V - \ell)R}$ for any $\ell$. 
We verify that $B \le (V-\ell)R$ if and only if $m \le V-\ell$, 
and the result follows. 
\end{IEEEproof}

From  Theorem~\ref{thm:upperBoundGen}, we can obtain an upper bound on the number of useful servers averaged over the number of fragments and the number of servers. 

\begin{rem}
\label{rem:UBSumNumUsefulServers}
Recall that for a completely utilizing $\alpha$-$(V,R)$ replication storage scheme $B/R = V/K = 1/\alpha$. 
When $B/R$ is an integer, 
summing up both sides of Eq.~\eqref{eq:UB}, 
and dividing both sums by the product $BV$, 
we obtain 
\EQN{
\label{eqn:NormalizedSumUsefulServersUB}
\frac{1}{BV}\sum_{\ell = 0}^{V-1}{N(I_{\ell})} 
\le 
1- \frac{(m+1)}{2V}.
}
This gives us a normalized upper bound on the sum of number of useful servers. 
\end{rem}
\begin{rem}
Note that this upper bound is true for all completely utilizing $\alpha$-$(V,R)$ replication storage schemes and all work-conserving scheduling policies.  
That is,
\EQ{
\sup_{\Phi \in \cS}\sup_{\Psi_\Phi}\frac{1}{BV}\sum_{\ell=0}^{V-1}N(I_\ell) \le 1 - \frac{(m+1)}{2V}. 
}
\end{rem}

\nix{Do not remove this nix
\begin{rem}
\label{rem:underutilizingUB}
Consider an underutilizing replication code stored on $B$ servers with storage capacity of $K$ fragments each, where  each fragment $v\in [V]$ is assumed to be replicated on $R_v$ servers, where $R_v \le R$ and $\frac{1}{V}\sum_{v \in [V]}R_v < R$.
Also, we may denote the number of fragments stored on each server $b$ as $K_b$ where $K_b \le K$. 
Since, the number of useful servers before the commencement of download can only be smaller than or equal to $B$ and $R_v \le R$ for all $v \in [V]$, following the exact steps as in above shows that the above given upperbound holds for this underutilizing replication code stored on $B$ servers with storage capacity $K$.
\end{rem}
}

\begin{rem}
\label{rem:underutilizingUB}
Consider an underutilizing $\alpha$-$(V,R)$ replication storage scheme where  each fragment $v\in [V]$ is assumed to be replicated on $R_v$ servers, where $R_v \le R$ and $\frac{1}{V}\sum_{v \in [V]}R_v < R$.
Since, the number of useful servers before the commencement of download can only be smaller than or equal to $B$ and $R_v \le R$ for all $v \in [V]$, following the exact steps as in above shows that the above given upperbound holds for an underutilizing $\alpha$-$(V,R)$ replication storage scheme.
\end{rem}

\subsection{Lower bound on $N(I_\ell)$}
Next, we proceed to find a lower bound for the number of useful servers $N(I_\ell)$ after $\ell$ downloads for a completely utilizing $\alpha$-$(V,R)$ storage scheme. 
To this end, we define two important properties of a storage scheme $\Phi \in \cS$. 
For servers $a,b \in [B]$ and fragments $v,w\in [V]$, 
we define the maximum overlap of fragment sets and occupancy sets as 
\begin{xalignat}{2}
\label{eqn:Overlap}
&\tau_M \triangleq \max_{a \neq b}\abs{S_a \cap S_b},&
&\lambda_M \triangleq \max_{v \neq w}\abs{\Phi_{v} \cap \Phi_{w}}.
\end{xalignat}

In the initial stages of download when the number of fragments downloaded is less than  $O(K^2)$,
the number of useful servers is primarily determined by the overlap between the fragment sets. 
More precisely, we have the following lower bound on $N(I_\ell)$.

\begin{thm}
\label{thm:LowerBoundNumUsefulLowerGen}
Consider a completely utilizing $\alpha$-$(V,R)$ replication storage scheme $\Phi$ with the maximum overlap of fragment sets $\tau_M$, 
as defined in Eq.~\eqref{eqn:Overlap}.  
Suppose the number of downloaded pieces satisfies 
\EQN{
\label{eq:lb-ell-gen}
\ell <  iK-i(i-1)\tau_M /2.
}
Then, for  $i \le \floor{K/\tau_M}+1$, 
the number of useful servers can be lower bounded as 
\EQN{
\label{eqn:UsefulLB1}
N(I_{\ell}) \ge B-i. 
} 
That is, there are at most $i$ servers that are no longer useful after $\ell$ downloads .
\end{thm}
\begin{IEEEproof}
We prove this by contradiction. 
We assume that the number of downloaded pieces $\ell$ satisfies Eq.~\eqref{eq:lb-ell-gen}, 
and the number of useful servers $N(I_\ell) < B-i$. 
We denote the number of useless servers $B-N(I_\ell) > i$. 
Then, there exists $i$ servers $\set{b_1, \dots, b_i} \subset [B]$, 
which are no longer useful after $\ell$ downloads.  This implies that the union of their fragment sets $\cup_{j=1}^iS_{b_j}$ is included in the downloaded fragment set $I_\ell$. 

Since any two servers can have at most $\tau_M$ file fragments in common, 
$j$th server has at least $(K-(j-1)\tau_M)$ fragments distinct from the fragments stored in first $(j-1)$ servers. 
Thus,
\EQ{
\ell \ge \abs{\cup_{j=1}^iS_{b_j}} \ge \sum_{j=0}^{i-1}(K-j\tau_M) = iK-i(i-1)\tau_M /2.
}
This contradicts our assumption and the result follows.
\end{IEEEproof}
\nix{
Please dot remove this nix
\bl{
\begin{note}
For an underutilizing $(VR',V)$ replication code stored on an $\alpha$-$B$ system as mentioned in Note~\ref{Note:underutilizingUB}, if  
the number of downloaded pieces satisfies 
\EQ{
\ell < \sum_{j \in U(I_{\ell})^c}K_j -i(i-1)\tau_M /2,
}
then the number of useful servers at $\ell$th download is greater than $B-i$.
\end{note}
}
}
By solving the quadratic inequality in Eq.~\eqref{eq:lb-ell-gen}, we can obtain the following result.

\begin{cor}
\label{cor:LowerBoundNumUsefulLowerGen}
Consider a completely utilizing $\alpha$-$(V,R)$ replication storage scheme $\Phi \in \cS$ with maximum fragment set overlap $\tau_M$.  
If the number of pieces downloaded  satisfies  $\ell < K(K+\tau_M) /2\tau_M$,
then the number of useful servers is at least 
\EQN{
\label{eq:Nil-lb-proj-des-1}
N(I_\ell) \ge B - \left(\frac{2K+\tau_M +\sqrt{(2K+\tau_M)^2-8\ell \tau_M}}{2\tau_M}\right).
}
\end{cor}

We can also give a lower bound on the number of useful servers when most of the fragments have been downloaded. 
In  this case the number of useful servers is determined by the overlap between occupancy sets.
By considering the largest possible overlap of the occupancy sets, we can show the following lower bound on $N(I_\ell)$.

\begin{thm} \label{thm:LowerBoundNumUsefulUpperGen}
Consider a completely utilizing $\alpha$-$(V,R)$ replication storage scheme $\Phi \in \cS$ with the maximum overlap of occupancy sets $\lambda_M$, as defined in Eq.~\eqref{eqn:Overlap}. 
Suppose that $\ell= V-i$ pieces have been downloaded, where $i \in \set{0, \dots, \floor{R/\lambda_M}+1}$. 
Then, the number of useful servers can be lower bounded as 
\EQN{
\label{eq:lb-Nil-last-K}
N(I_\ell) \ge iR-i(i-1)\lambda_M/2.
}
\end{thm}
\begin{IEEEproof} 
After $\ell$ downloads, 
the set of downloaded fragments is $I_\ell$, 
and the set of remaining fragments is $[V]\setminus I_\ell$. 
Since $\ell = V-i$, we denote the set of remaining fragments as $\set{w_1, \dots, w_i}$. 
Recall that the set of useful servers is the union of servers storing the remaining fragments $U(I_\ell) = \cup_{v \notin I_\ell}\Phi_v = \cup_{j=1}^i\Phi_{w_j}$. 
We can write the number of useful servers as \EQ{
N(I_\ell) 
= \abs{\cup_{j=1}^i\Phi_{w_j}}
= \sum_{j=1}^i(\abs{\Phi_{w_j}}- \abs{\cup_{r=1}^{j-1}(\Phi_{w_j}\cap\Phi_{w_r})}).
}
Since any two occupancy sets can have at most $\lambda_M$ servers in common, 
we get 
\EQ{
N(I_\ell) \ge \sum_{j=1}^{i}(R - (j-1)\lambda_M),
}
and the result follows. 
\end{IEEEproof}
\nix{
Please do not remove this nix.
\bl{
\begin{note}
For an underutilizing $(VR',V)$ replication code stored on an $\alpha$-$B$ system as mentioned in Note~\ref{Note:underutilizingUB}, if the number of pieces downloaded, $\ell = V-i$,
the number of useful servers satisfies 
\EQ{
N(I_{\ell}) \ge  \sum_{j \in I_{\ell}^c}R_j -i(i-1) \lambda_M /2.
}
\end{note}
}
}
For any completely utilizing $\alpha$-$(V,R)$ replication storage scheme, 
we observe that the proposed lower bounds on the number of useful servers depends on the maximum overlap fragment set overlap $\tau_M$ and the maximum overlap of the occupancy sets $\lambda_M$. 
This bound holds for any work-conserving scheduling policy $\Psi_\Phi$ restricted to this storage scheme. 
Replication storage schemes with small values for the overlap parameters $\tau_M$ and $\lambda_M$ will maximize the derived lower bounds. 
Motivated by this fact, we introduce certain special replication storage schemes with desirable overlap parameters. 
When the maximum overlap of fragments $\tau_M$ is $1$, we have the following lower bound on the number of useful servers.
\begin{lem}
\label{lem:LowerBoundBIBD}
For a completely utilizing $\alpha$-$(V,R)$ replication storage scheme
with the maximum overlap of fragments $\tau_M$ set to be $1$,
the number of useful servers $N(I_\ell)$ satisfies
\EQ{
N(I_{\ell}) \ge N(I_{\ell-1}) - \floor{\frac{\ell-1}{K-1}}.
}
\end{lem}
\begin{IEEEproof}
We denote the set of servers that turn useless after $\ell$th download by $G_\ell \triangleq U(I_{\ell-1})\setminus U(I_\ell)$, 
and its cardinality by $\abs{G_\ell} = N(I_{\ell-1}) -  N(I_\ell)$. 
Each server $b \in G_\ell$ has the $\ell$th downloaded fragment $v_\ell$, and all its remaining fragments have been downloaded in first $(\ell-1)$ downloads. 
Further, since the maximum overlap of fragment sets for these storage schemes is $\tau_M = 1$, 
it follows that the sets $(S_b\setminus \set{v_\ell}: b \in G_\ell)$ are disjoint.  
That is, we can write 
\EQ{
\cup_{b \in G_\ell}(S_b\setminus\set{v_\ell}) \subseteq I_{\ell-1}.
}
Since $\abs{I_{\ell-1}}=\ell-1$ and $\abs{S_b\setminus\set{v_\ell}} = K-1$ for all $b \in G_\ell$, we obtain the result. 
\end{IEEEproof}

	\section{Deterministic placement schemes for replication codes} 
\label{sec:DesignBasedStorage}
In this section we shift our attention to Problem~\ref{prob:OptPlace}.  
Recall from the discussion in Section~\ref{sec:ProblemForm}, 
given the work-conserving scheduling policy, 
one can find the optimal storage scheme using a brute-force search among all the storage schemes. 
As the brute-force is computationally expensive,  and we need the knowledge of the work-conserving scheduling policy, there is no obvious efficient  solution to this problem. 
In particular, the dependency on the work-conserving scheduling policy  makes Problem~\ref{prob:OptPlace} difficult to solve directly.  

In Problem~\ref{prob:OptPlace}, we see that the optimal storage scheme maximizes the mean number of useful servers aggregated over all fragment downloads. 
Theorems~\ref{thm:LowerBoundNumUsefulLowerGen}~and~\ref{thm:LowerBoundNumUsefulUpperGen} provide deterministic lower bounds on the number of useful servers after each fragment download, 
for any work-conserving scheduling policy. 
This suggests that one could search for the storage schemes that maximize the lower bound on the number of useful servers. 
This allow us to systematically approach Problem~\ref{prob:OptPlace} without knowledge of any work-conserving scheduling policy. 
The resulting storage schemes are empirically shown to have good performance. 

Lower bounds on the number of useful servers are decreasing in the overlap parameters $\tau_M$ and $\lambda_M$. 
This suggests that keeping the overlap between the servers small, we can make the servers useful for a longer time. 
We see in Theorem~\ref{thm:LowerBoundNumUsefulLowerGen} that the maximum overlap $\tau_M$ between fragment sets affects the initial stage of downloading the file.
Further, we observe in Theorem~\ref{thm:LowerBoundNumUsefulUpperGen} that the maximum overlap $\lambda_M$ between occupancy sets affects
the final stages of downloading the file. 

Using this intuition we propose some algebraic constructions for storage schemes.
We establish a connection between storage schemes and combinatorial designs. 
This connection allows us to systematically construct new storage schemes for a given $\alpha$-$B$ system.  
We then study the performance of a particular class of storage schemes constructed from 
a combinatorial design called projective plane. 
Numerical studies are reported in Section~\ref{sec:sims}.

\subsection{Storage schemes from combinatorial designs}

In this section, we establish a correspondence between storage schemes and designs. 
Then using this correspondence we propose storage schemes from designs, 
with desirable overlap parameters $\tau_M$ and $\lambda_M$. 
\begin{defn}[Design]
A design is a pair $(\cP, \cB)$ satisfying the following conditions:
\begin{compactenum}[D1)]
\item $\cP$ is a set of elements called points. 
\item $\cB$ is a collection of nonempty subsets of $\cP$ called blocks.
\end{compactenum}
\end{defn}

\begin{thm}
\label{th:codes-designs}
Every completely utilizing $\alpha$-$(V,R)$ replication storage scheme corresponds to a design $([V],\cB)$  
where $\abs{\cB} = B = \frac{R}{\alpha}$ 
and 
\EQ{
\cB = \set{ S_b \subseteq [V]:   b\in [B]}. 
%\cB = \red{(S_b \subseteq [V]:   b\in [B]).} 
}
Conversely, every design $(\cP,\cB)$ leads to a replication storage scheme with $\abs{\cP}$ fragments stored on $B=\abs{\cB}$ servers, 
where $b$th server is storing the fragments indexed by the $b$th block of $\cB$. 

In addition, if every block has the same size $K$, 
and every point occurs $R$ times, then we obtain an $\alpha$-$(\abs{\cP}, R)$ replication storage scheme where $\alpha = R/\abs{\cB}$.
\end{thm}
\begin{IEEEproof}
Given a completely utilizing $\alpha$-$(V, R)$ replication storage scheme we can form a design as follows:
i) identify all the fragments $[V]$ with the point set of the design $\cP$, 
ii) identify $S_b$ the set of fragment replicas at each server with a block. 
Thus the collection of blocks $\cB= \{S_b \subseteq [V]: b\in [B]\}$.

Conversely, suppose we are given a design $(\cP, \cB)$.
Construct a storage scheme as follows. 
Divide the file into $V = \abs{\cP}$ fragments. 
Then there is a one to one correspondence between the points and the integers in $[V]$. 
For each block in $\cB$, we associate a server which stores the fragments corresponding to
the points in that block. 
Since each block can be associated to a server, 
we have $B=|\cB|$ servers. 
If all blocks are of the same size $K$ and every point occurs $R$ times, 
then this is a completely utilizing $\alpha$-$(V, R)$ replication storage scheme for $\alpha = \frac{K}{V} = \frac{R}{B}$.
\end{IEEEproof}

The class of designs that are suitable for completely utilizing $\alpha$-$(V,R)$ replication storage schemes, 
where all the servers have the same storage capacity, 
are the so-called  $t$-designs.
A design $(\cP, \cB)$ is said to be a $t$-design with parameters $t$-$(V, K, \lambda)$ if
\begin{compactenum}
\item there are $V$ points in $\cP$, %i.e. $|\cP|=V$
\item every block in $\cB$ contains exactly $K$ points, %i.e. $\abs{S_b} = K$ for all $b \in \cB$, and 
\item every $t$-subset of $\cP$ is contained exactly in $\lambda$ blocks of $\cB$. 
\end{compactenum}

The following result is well known from design theory and can be found in any standard textbook, see
for instance~\cite{stinson03}. 
For completeness, a short proof is provided in the Appendix~\ref{app:designs}. 

\begin{prop}[\cite{stinson03}]
\label{prop:design prelims}
	For every
	$t$-$(V, K, \lambda)$ design,
	the following conservation laws hold true,
	\eqn{
		\label{eq:des-req-1}
		BK &=VR,\\
		\label{eq:des-req-2}
		B\binom{K}{t} &= \lambda  \binom{V}{t},
	}
	where $B$ is the number of blocks and $R$ is the number blocks containing any point. 
\end{prop}

From the one to one correspondence between designs and replication storage schemes shown in Theorem~\ref{th:codes-designs}, 
it follows that the set of blocks in which a point $p$ appears is precisely the occupancy set of the fragment corresponding to point $p$.
Further,  the total number of  blocks that contain the  point $p$ is the  replication factor $R$
of the corresponding fragment. 
In our setting, the points are the fragments, and the blocks are the set of fragments on each server. 
Table~\ref{tab:designs2codes} summarizes the mapping between design and storage parameters.
\begin{table}[H]
    \centering
        \caption{Correspondence between designs and storage codes}
		\begin{tabular}{ l| l }
			\hline \hline
			\multicolumn{2}{c}{$t$-$(V, K, \lambda)$  designs to codes}\\
			\hline
			Design parameter & Storage parameter  \\	\hline
			$\cP$: Points  & $[V]$: File fragments\\ \hline
			$\cB$: Blocks & $(S_b: b\in [B])$: Fragment sets at servers \\ \hline
			$|\cP|$:Number of points &  $V$:Number of file fragments  \\  \hline
			$|\cB|$: Number of blocks  & $B$: Number of servers   \\ \hline
			$K$: Size of each block & $K$: Storage capacity at each server  \\ \hline
			$R$: Replication factor  for each point & $R$: Replication factor  for each fragment \\ \hline
			\hline
		\end{tabular}
    \label{tab:designs2codes}
\end{table}

In practical systems, typically, the number of servers and the storage capacity at each server is fixed. 
In addition,  Eq.~\eqref{eq:des-req-1} tells us that among $V$ and $R$, 
only one can be chosen independently.
That is, among the two design parameters, the number of pieces in to which file is fragmented and the replication factor, choosing one determines the other one.
Similarly, having chosen $V$, the parameter $\lambda$ is completely known for any $t$-design.

In general, it is an open question whether a  $t$-design with a given set of parameters exists and if it exists how to construct that design. 
For some specific parameters, there are explicit constructions of designs. 
A popular and well studied case is where $t=2$. 
These designs are called balanced incomplete block designs (BIBDs).
Then, from Eq.~\eqref{eq:des-req-2}, we have 
\EQN{
	\label{eq: num_blks}
	B K (K-1) = \lambda V (V-1),
}
and using Eq.~\eqref{eq:des-req-1} we obtain
\EQN{
	\lambda = \frac{R (K-1)}{(V-1)}. \label{eq:lambda-bibd}
}	

A special case of  BIBDs of interest is where the number of points is equal to the number of blocks. 
In other words, the  number of servers is equal to the number of fragments. 
Such BIBDs are also called symmetric BIBDs. 
Then, from Eq.~\eqref{eq:des-req-1}, we note that the replication factor is equal to the memory at each server. 
Further, we also have the property that any two distinct blocks intersect in $\lambda$ points, see~\cite[Theorem~2.2]{stinson03}. 
An important symmetric design is the projective plane for which explicit constructions are known. 

Another well known 2-design for which explicit construction is known is the affine plane, 
and can be obtained from a projective plane. 
The memory requirements at each server scale as $O(\sqrt{B})$ for storage schemes derived from projective planes (and affine planes) with $B$ servers. 
If we are interested in storage schemes with a fixed size of memory, then we could consider constructing a storage scheme from a Steiner triple systems~\cite{stinson03}. 
They are also 2-designs, but with the block size fixed at $K=3$.
In the following table we summarize the storage schemes from the designs we discussed above.
\begin{table}[H]
    \centering
            \caption{Parameters of various completely utilizing $\alpha$-$(V,R)$ replication storage schemes from designs.
            }
 		\begin{tabular}{ l|l| l }
			\hline \hline
%			\multicolumn{2}{c}{Designs to codes}\\
%			\hline
			 $t$-design & Parameters& $\alpha$-$(V, R)$ system   \\ 
			\hline
					General  & $t$-$(V, K, \lambda)$& $\frac{K}{V}$-$(V, R)$\\ 
			$t$-design &  & $B=\lambda\binom{V}{t}/\binom{K}{t}$  \\  
			 &  &  $R=\lambda\binom{V-1}{t-1}/\binom{K-1}{t-1}$\\  \hline
			BIBD  &  2-$(V, K, \lambda)$ & {$\frac{K}{V}$-$(V, R)$} \\ 
		     && $B=VK/R$, $R=\lambda (V-1)/(K-1)$\\ \hline
			Symmetric  &  2-$(V, K, \lambda)$ &$\frac{K}{V}$-$(V, K)$ \\ 
			 BIBD& & $B=V$, $R=\lambda (V-1)/(K-1)$ \\ \hline
			Projective  & 2-$(n, q+1,1)$,&{$\frac{q+1}{n}$-$(n, q+1)$}\\ 
			 plane& $n=q^2+q+1$ & $B=n$, $R=q+1$\\ 
			  & $q=p^m$, $p$ prime & $K=q+1$\\\hline
			Affine  &  2-$(q^2, q, 1)$ & {$\frac{q}{q^2}$-$(q^2,q+1)$} \\ 
			plane &$q=p^m$, $p$ prime &   $B=q^2+q$, $K=q$\\ 
						    \hline
						    Steiner triple & $2$-$(V, 3, 1)$& {$\frac{3}{V}$-$(3V/R, V, R)$}\\
						    system &$V\equiv 1, 3 \bmod 6$&$R=(V-1)/2$\\
			\hline
		\end{tabular}
    \label{tab:bvrk-params}
\end{table}
As can be seen from the Table~\ref{tab:bvrk-params}, 
there is a flexibility in the choice of the design parameters allowing us to construct storage schemes  for various $ (B, K) $ systems. Additional storage schemes can be constructed using  results from design theory. 
For instance a $t$-$(V, K, \lambda)$ design implies the existence of $(t-i)$-$(V-i, K-i, \lambda)$
design for $i<t$, see \cite[Theorem~9.2]{stinson03}. 

\subsection{Storage schemes from projective planes}

As mentioned earlier Theorems~\ref{thm:LowerBoundNumUsefulLowerGen}~and~\ref{thm:LowerBoundNumUsefulUpperGen} motivate us to  
construct storage schemes where  $\lambda_M$ 
and $\tau_M$ are small. 
Since they are both nonnegative, one might try to make them both zero. 
However, $\lambda_M=0$   implies that $K=1$ which has been studied extensively.
Similarly,  $\tau_M=0$  implies  that $R=1$ in which case there is no redundancy. 
For these reasons, we do not study  these two cases in this paper. 
The next possible choice would be $\lambda_M=1$ and $\tau_M=1$.
Note, that $\lambda_M=1$ implies that the maximum occupancy overlap is less than or 
equal to one.
For simplicity, we consider the symmetric case, where for any two distinct fragments $u$, $v$
we have $|\Phi_u\cap \Phi_v| = 1$. 
Such a storage scheme immediately leads us to  Steiner systems which are BIBDs with $\lambda=1$. 
This motivates the study of storage schemes from such $2$-designs. 
If we similarly restrict that the fragment sets  also satisfy a similar overlap property, 
i.e. $|S_a\cap S_b|=1$ for distinct blocks $a$, $b$, then
such a BIBD must also be symmetric in that number of blocks is identical to the 
number of fragments \cite[Corollary~2.5]{stinson03}. 
A well studied class of symmetric BIBDs is that of projective planes. 
In this section, we study storage schemes from projective planes. 
If we relax the constraint that two blocks do not necessarily intersect in $\lambda_M$ points, then 
also it is possible to construct a storage scheme form a 2-design with 
$\lambda_M=\tau_M=1$. 
Specifically, the affine planes lead to storage schemes with these parameters. 
In the rest of the section, we review the construction of projective planes and affine planes
and use them to construct storage schemes. 

We now briefly review the construction of a projective plane from $\F_q^3$ where $q$ is the power of a prime
and $\F_q$ denotes a finite field with $q$ elements, see \cite{stinson03} for more details. 
Consider the set of $3$-tuples in $\F_q^3$. 
Let $\fV_i$ be the set of all $i$-dimensional subspaces of $\F_q^3$. 
The projective plane is formed by taking the collection of points to be $\fV_1$ and 
collection of blocks to be $\fV_2$ giving us a 
2-$(q^2+q+1,q+1,1)$ design. 
Let $U\in \fV_2$, then we can associate to it a block as follows.
\begin{eqnarray}
B_U = \set{ V\in \fV_1 \mid V\subset U }.
\end{eqnarray}
The collection of blocks is then given by 
\begin{eqnarray}
\cB = \set{ B_U\mid U\in \fV_2}.
\end{eqnarray}

Since scalar multiples of a nonzero vector $(x_1, x_2, x_3)$ are in the same vector space, we 
can see that there are $(q^3-1)/(q-1)=q^2+q+1$ vectors in $\F_q^3$
 which are not scalar multiples of each other. 
Thus there are $q^2+q+1$ vector spaces in $\fV_1$.
For each 2-dimensional vector space we associate a 1-dimensional orthogonal vector space.
This is unique since $(V^\perp)^\perp=V$.
Therefore, the number of 2-dimensional vector spaces $|\fV_2|$ is equal to the number of 1-dimensional vector spaces $|\fV_1|$. 
In each 2-dimensional vector space $U$, there are $q^2-1$ nonzero vectors. 
Each nonzero vector $u$ generates a 1-dimensional vector space. 
The nonzero nonscalar multiples of $u$ also generate the same vector space.
Since there are $q-1$ nonzero scalar multiples of any vector, 
these $q^2-1$  nonzero vectors generate  
$q+1$ distinct 1-dimensional vector subspaces in $U$.
Thus  there are $(q+1)$ distinct 1-dimensional subspaces in each 2-dimensional vector space. 
This implies that each block of the projective plane has $q+1$ points. 

Any two distinct one-dimensional spaces $V, V' \in \fV_1$ must be generated by distinct vectors
$v, v'$ respectively. 
If $V, V'$ are both subspaces of a two-dimensional subspace $U\in \fV_2$, then $v, v'$ are also in 
$U$.
Thus $U$ must be generated by $v, v'$. 
Any other 2-dimensional vector space $W$ that contains $V, V'$ must be identical to $U$.
Thus two distinct one-dimensional vector spaces can occur only in one two-dimensional vector space giving $\lambda=1$.
From these designs we obtain an $\alpha$-$(q^2+q+1, q+1)$ replication storage scheme, where $\alpha = \frac{(q+1)}{(q^2+q+1)}$. 
A $\frac{3}{7}$-$(7, 3)$ replication storage scheme constructed from a projective plane is shown in Table~\ref{tab:design-7733}.

\begin{table}[htbp]	
	\begin{center}
		\begin{tabular}{c|c|c|c|c|c|c|c}
			\hline \hline
			\multicolumn{8}{c}{Placement of pieces}\\
			\hline
			Server   & 1 & 2 & 3 & 4 & 5 & 6 & 7 \\ \hline
			\nix{
			Layer 1  & 1& 3 & 1 & 1 & 2 & 3 & 2 \\ 
			Layer 2  & 2 & 4 & 5 & 4 & 5 & 6 & 4 \\ 
			Layer 3  & 3 & 5 & 6 & 7 & 7 & 7 & 6 \\ \hline
			}
		$S_b$	&1,2,3 & 3,4,5&1,5,6& 1,4,7&2,5,7&3,6,7&2,4,6\\
			\hline
		\end{tabular}
	\end{center}
	\caption{A $\frac{3}{7}$-$( 7, 3)$ replication storage scheme based on a projective plane.}
	\label{tab:design-7733}
\end{table}

Given a design we can often construct other designs from it. 
For instance, given a projective plane, we can construct another design called the affine plane.
In general, to construct an affine plane, we start with a projective plane. 
We take any 2-dimensional vector space in $\fV_2$ and remove the points of this block from the set of points and all the remaining blocks. 
Since each block contains $q+1$ points, 
the number of blocks is reduced by unity, and the number of points is reduced by $q+1$. 
As there are $q^2+q+1$ blocks (and points) originally, the residual points are $q^2$ in number and the remaining blocks are $q^2+q$ in number. 
Every remaining block has an overlap of $\lambda=1$ with the deleted block, thus 
each residual block contains exactly $q$ points. 
\nix{[Explanation for why only one point is reduced in remaining blocks.]}
Finally, note the residual blocks are a subset of the blocks of the projective plane. 
Therefore, any pair of residual points occur in exactly one block as they did in the
projective plane. 
This gives us a 2-$(q^2, q, 1)$ design.
From this design we can construct a 
$\frac{1}{q}$-$(q^2, q+1)$ replication storage scheme. 

\subsection{Analysis of storage schemes derived from projective plane designs}

\begin{rem} 
Recall that the overlap parameters $\tau_M = \lambda_M = 1$ for any $\alpha$-$(V,R)$ replication storage scheme derived from a $2$-$(q^2 + q + 1,q+1,1)$ design. 
Therefore, from  Corollary~\ref{cor:LowerBoundNumUsefulLowerGen},  Theorem~\ref{thm:LowerBoundNumUsefulUpperGen}, and Lemma~\ref{lem:LowerBoundBIBD}, 
we can lower bound the number of useful servers for number of downloads $\ell \in \set{0, \dots, V}$, as 

\eqn{
\label{eq:LB-DesignBased}
N(I_\ell) \ge \begin{cases}
B - \big(\frac{2K+1 +\sqrt{(2K+1)^2-8\ell}}{2}\big), &\ell \le \frac{K(K+1)}{2},\\
(V-\ell)R-\frac{(V-\ell)(V-\ell-1)}{2}, &\ell \ge V-K-1,\\
N(I_{\ell-1}) - \floor{\frac{\ell-1}{K-1}},&\text{else}.
\end{cases}
}\label{}
\end{rem}

\subsection{Cyclic shift based storage} 
\label{subsec:Cyclic}
We conclude this section by considering a simple $\alpha$-$(V, R)$ storage scheme.
The set of servers storing replicas of a fragment $v \in [V]$ are cyclically  shifted by $v$.
More precisely, the occupancy set of a fragment $v$ is given by 
\EQN{
\label{eqn:OccupancySetCyclic}
\Phi_{v+R-1} = \set{v, v+1, \cdots, v+R-1},
}
where the addition is modulo $B$ whenever the sum exceeds $B$. 
We observe that $\abs{\Phi_{v} \cap \Phi_{v+j}} = \abs{\Phi_v \cap \Phi_{v-j}} = R - j$ for $j \in [R-1]$, $v \in [V]$ where the addition and subtraction is modulo $V$ whenever the sum exceeds $V$ or difference go below $1$.
Thus, the maximum overlap between the occupancy sets $\lambda_M = R-1$.

In addition, we can write the fragment set at server $b \in [B]$ for this storage scheme as 
\EQN{
\label{eqn:FragmentSetCyclic}
S_b = \set{b,b+1, \dots, b+K-1},
}
where the addition is modulo $V$ whenever the sum exceeds $V$. 
Here, we observe that $\abs{S_{b+j} \cap S_b} = \abs{S_{b-j} \cap S_b} = K-j$ for $j \in [K-1]$, $b \in [B]$ where the addition and subtraction is modulo $B$ whenever the sum exceeds $B$ or difference go below $1$.
Thus, the maximum overlap between the servers $\tau_M = K-1$.
An example is given in Table~\ref{tab:cyclic-matching}.
\begin{table}[htbp]
	\begin{center}
		\begin{tabular}{c|c|c|c|c|c|c|c}
			\hline \hline
			\multicolumn{8}{c}{Placement of pieces}\\
			\hline
			Server   & 1 & 2 & 3 & 4 & 5 & 6 & 7 \\ \hline
			\nix{
			Layer 1  & 1 & 2 & 3 & 4 & 5 & 6 & 7 \\ 
			Layer 2  & 2 & 3 & 4 & 5 & 6 & 7 & 1 \\ 
			Layer 3  & 3 & 4 & 5 & 6 & 7 & 1 & 2 \\ \hline
			}
			$S_b$&1,2,3&2,3,4&3,4,5&4,5,6&5,6,7&6,7,1&7,1,2\\
			\hline
		\end{tabular}
	\end{center}
	\caption{A $\frac{3}{7}$-$(7,3)$ replication storage schemes based on cyclic shift. }
	\label{tab:cyclic-matching}
\end{table}	

While the maximum overlap parameters for this scheme are $\tau_M= K-1$ and $\lambda_M=R-1$, 
the overlap parameters have a wide spread.  
Any server will have a fragment set overlap $i$, where $ i \in [K-1]$, with exactly two other servers and will have zero overlap with the remaining $\max(0, B-2K+1)$ servers.
Similarly, the occupancy set of each fragment will have an overlap of $j$, where $ j \in [R-1]$, with the occupancy sets of exactly two other fragments and will have zero overlap with the occupancy sets of remaining $\max(0,V-2R+1)$ fragments.

From the bounds in Theorems~\ref{thm:LowerBoundNumUsefulLowerGen}~and~\ref{thm:LowerBoundNumUsefulUpperGen},  we expect that these schemes  should perform somewhat poorly compared to the 
schemes presented previously, as we will corroborate in Section~\ref{sec:sims}. 
If $B=V$, the fragments in the $i$th locations of the servers can be arranged to be a permutation of $[V]$ giving a natural 
nonadaptive scheduling policy.
Later, in Section~\ref{sec:sims} we will study the performance of this scheme  along with a adaptive scheduling policy to highlight the importance of scheduling. 
The worst case overlap parameters for cyclic shift based storage schemes are very large,  
and this is reflected in their large mean download time with nonadaptive scheduling. 
However, the performance is not only affected by 
$\lambda_M$ and $\tau_M$ 
but also the spread of the occupancy set overlaps and fragment set overlaps. 
Taking advantage of this fact we can improve the performance of this scheme with 
adaptive scheduling. 
Specifically, the adaptive schedulng algorithms exploit the spread of the overlap parameters, 
to drive the system state towards good residual fragment sets with small overlap. 

	\section{Work-conserving Scheduling}
\label{sec:Scheduling}

Recall that our original Problem~\ref{prob:MeanServiceCompletionTime} was broken down into two interrelated Problems~\ref{prob:OptSched}~and~\ref{prob:OptPlace}.
We addressed Problem~\ref{prob:OptPlace} by obviating the need for the knowledge of an optimal scheduling policy. 
In order to complete our solution for Problem~\ref{prob:MeanServiceCompletionTime}, 
we need to combine it with a solution for Problem~\ref{prob:OptSched}.
As already discussed in Section~\ref{sec:ProblemForm},  Problem~\ref{prob:OptSched} can be reformulated as an MDP, given a completely utilizing $\alpha$-$(V, R)$ storage scheme. 
In view of the complexity of the MDP, we propose two classes of efficient suboptimal work-conserving scheduling algorithms.
One class of algorithms are nonadaptive while the other are adaptive.
The nonadaptive algorithms fix the schedule at each server ahead of the file download.
In adaptive algorithms, the schedule of remaining fragments to be downloaded at each server is causally aware of the sequence of fragment downloads.
We show that both classes of algorithms offer good performance through numerical studies.
As will be seen in Section~\ref{sec:sims}, adaptive algorithms can give a better performance than the nonadaptive ones. 
Adaptive work-conserving scheduling provides more flexibility in the download process which can be exploited to reduce the mean download time. 
Nonadaptive scheduling policies maybe preferred in some cases, 
when adaptive fragment selection incurs non-negligible delay. 

\subsection{Nonadaptive work-conserving scheduling}

Given a completely utilizing $\alpha$-$(V,R)$ replication storage scheme $\Phi$, 
we can find the fragment set $S_b$ at each server $b \in [B]$. 
In the nonadaptive case, the scheduling decisions are embedded in the placement order $\pi_\Phi^b: [K] \to S_b$ of fragment set at each server $b \in [B]$, 
and is fixed prior to the commencement of download. 
A nonadaptive work-conserving scheduling is defined by the collection of placement order $\pi_\Phi \triangleq (\pi_\Phi^b: b \in [B])$. 
Given the set of downloaded fragments $I_\ell$ after $\ell$ downloads, 
the fragment to be downloaded from a server $b$ is the first residual fragment stored at this server in the order of placement. 
This fragment is denoted by $\pi_\Phi^b(k^b_{\ell+1})$, 
where 
\EQ{
k^b_{\ell+1} \triangleq \inf\set{k \in [K]: \pi_{b}(k) \notin I_\ell},\quad b \in U(I_\ell).
} 
A placement order $\pi$ induces a scheduling policy $\Psi$ such that 
$\Psi_\Phi(I_\ell)(b) = \pi_\Phi^b(k^b_{\ell+1})$. 
Finding a nonadaptive work-conserving scheduling policy is equivalent to finding a placement order.
Therefore,  given an $\alpha$-$(V,R)$ replication storage scheme $\Phi \in \cS$, 
the optimal nonadaptive work-conserving scheduling policy is given by
\EQ{
\pi^\dagger(\Phi) = \arg\max_{\pi(\Phi)}\frac{1}{V}\sum_{\ell=0}^{V-1}\E{N(I_\ell)}.
}
One way to find the optimal nonadaptive work-conserving scheduling policy is to search over all possible placement orderings. 
This search is highly computationally intensive, 
and it is not clear how to efficiently find the optimal nonadaptive work-conserving scheduling policy. 
As such, we discuss heuristic nonadaptive work-conserving scheduling policies in Section~\ref{sec:sims} that attempt to maximize the mean number of useful servers, aggregated over all downloads. 

\subsection{Adaptive work-conserving scheduling}

Recall that, 
for a fixed completely utilizing $\alpha$-$(V,R)$ replication storage scheme $\Phi$, 
the corresponding adaptive scheduling policy is a map  $\Psi_\Phi: 2^{[V]} \to [V]^{[B]}$. 
In particular, the policy schedules a residual fragment $\Psi_\Phi(I_\ell)(b) = w^b_{\ell+1} \in S_b^\ell$ on each useful server $b \in U(I_\ell)$, after $\ell$ downloads. 
That is,
\EQ{
\Psi_\Phi(I_\ell): b \mapsto S_b^\ell, \text{ for all } b \in U(I_\ell), \text{ all } \ell \in \set{0, \dots, V-1}. 
}

We first show that the evolution of downloaded fragments can be modelled as a Markov chain for any work-conserving scheduling policy $\Psi_\Phi$ given a fixed completely utilizing $\alpha$-$(V,R)$ replication storage scheme $\Phi$. 
We then pose optimal dynamic scheduling problem defined in Problem~\ref{prob:OptSched} as an MDP.

Let $X_t \subseteq [V]$ be the set of downloaded fragments at time $t$. 
We can write $\ell$th download instants in terms of the process $X \triangleq (X_t \in 2^{[V]}: t \in \Z_+)$ as 
\EQ{
D_\ell = \inf\set{t > 0: \abs{X_t} = \ell}.
} 
At $\ell$th download instant $D_\ell$, the set of downloaded fragments is $I_\ell = X_{D_\ell}$. 
Further, the set of useful servers at this instant is $U(I_\ell)$. 
At time $D_\ell$, 
the fragment scheduled on any useful server $b \in U(I_\ell)$ is the scheduling decision $\Psi_\Phi(I_\ell)(b) \in S_b^\ell$. 
Since the process $X$ is piecewise constant and only changes at decision epochs, 
we are interested in the associated discrete time process sampled at the decision epochs $\set{D_0, D_{1}, \dots, D_{V-1}}$. 
Defining $Y_\ell \triangleq X_{D_\ell}$ for all $\ell \in \set{0, \dots, V-1}$, 
we can write the sampled process as $Y \triangleq (Y_\ell: \ell \in \set{0, \dots, V-1})$. 
\begin{lem} 
\label{lem:SchedulingCTMC}
For a fixed completely utilizing $\alpha$-$(V,R)$ replication storage scheme $\Phi$, scheduling policy $\Psi_\Phi$, and \iid exponential fragment download times, 
the continuous time process $X = (X_t \in 2^{[V]}: t \in \Z_+)$ is Markov. 
Hence, the associated sampled process $Y$ is a discrete time Markov chain with the transition probabilities given by 
\EQN{
\label{eqn:ProbDownloadFragment}
p_{I_\ell, I_{\ell}\cup\set{v}} = \frac{1}{N(I_\ell)}\sum_{b \in U(I_\ell)}\SetIn{\Psi_\Phi(I_\ell)(b)=v},\quad v \notin I_\ell. 
}
\end{lem}
\begin{IEEEproof}
Refer Appendix~\ref{app:SchedulingCTMCProof}.
\end{IEEEproof}
\nix{
\begin{rem} 
Since $X$ is a continuous time Markov process with finite states, 
the sampled process $Y$ at the instant of fragment download times is also Markov, 
with the following transition probabilities 
\EQ{
p_{I_\ell, I_{\ell}\cup\set{v}} = 
 \frac{1}{N(I_\ell)}\sum_{b \in U(I_\ell)}\SetIn{\Psi_\Phi(I_\ell)(b)=v},\quad v \in [V]\setminus I_\ell. 
}
\end{rem}
}
\subsubsection{MDP formulation}
Since we have already shown that the set of downloaded fragments evolves as a discrete-time Markov chain at the decision epochs, 
we can reformulate Problem~\ref{prob:OptSched} as an MDP. 
Recall that the objective function $\frac{1}{V}\sum_{\ell=0}^{V-1}\E{N(I_\ell)}$ is additive over the downloads. 
Therefore, we can consider this to be a finite MDP with $V$ stages, 
where the reward in stage~$\ell$ is 
\EQN{
\label{eqn:PerStageReward}
r_\ell(I_\ell) \triangleq \frac{N(I_\ell)}{V},\quad \ell \in \set{0, \dots, V-1}.
}
 Given a fixed completely utilizing $\alpha$-$(V,R)$ replication storage scheme $\Phi$, 
our goal is to find the optimal work-conserving scheduling policy $\Psi_\Phi$ such that the aggregate reward is maximized over the finite time horizon, i.e. 
\EQN{
\label{eqn:MDP}
\Psi^\ast_\Phi = \arg\max\E{\sum_{\ell=0}^{V-1}r_\ell(I_\ell)}.
}

A very powerful idea to solve an MDP is to use principle of optimality~\cite{Puterman1994Wiley}. 
This principle implies the following property for any optimal policy. 
\begin{quote} 
Whatever the current state and decision, the remaining decisions
must constitute an optimal policy with regard to the state resulting from the current decision. 
\end{quote}
Since the rewards are additive for our current problem, we can define a reward-to-go function $u_{\ell+1}(\Psi_\Phi(I_\ell)) = \E{\sum_{j=\ell+1}^{V-1}r_j(I_j)}$ at the $\ell$th download time instant $D_{\ell}$, of the current state $I_\ell$ and the decision rule $\Psi_\Phi(I_\ell)$. 
We can re-write this reward-to-go function as
\EQ{
u_{\ell+1}(\Psi_\Phi(I_\ell))  
= \sum_{v \notin I_\ell} p_{I_\ell, I_\ell\cup\set{v}} \Big[r_{\ell+1}(I_\ell\cup\set{v}) + u_{\ell+2}(\Psi_\Phi(I_\ell\cup\set{v}))\Big].
}
Then, the optimal reward-to-go function from $(\ell+1)$th stage is given by 
\EQ{
u^*_{\ell+1}(I_\ell) = \max_{\Psi_\Phi}u_{\ell+1}(\Psi_\Phi(I_\ell)),\quad \ell \in \set{0, \dots, V-1}.
}

This is the well known \emph{Bellman's optimality equations} and it follows from \cite[Theorem 4.3.3]{Puterman1994Wiley} that
\EQ{
u^*_0(I_{0}) = \max_{\Psi_\Phi} \E{\sum_{\ell=0}^{V-1}r_\ell(I_\ell)}
}
and the optimal work-conserving scheduling policy is the one that achieves the optimal reward-to-go function 
i.e.
\EQ{
\Psi^\ast_\Phi =\arg\max_{\Psi_\Phi} u_{0}(\Psi_\Phi(I_0)).
}
In order to solve this optimal scheduling decision we can use the standard backward induction algorithm\cite{Puterman1994Wiley}. 
For completeness, we have provided the backward induction algorithm tailored to our problem in Appendix~\ref{app:BackwardInduction}. 
The computational complexity of the backward induction algorithm to solve the optimal scheduling problem grows exponentially with the number of file fragments as discussed in Appendix~\ref{subsec:CompComplexMDP}. 

\subsubsection{Greedy scheduler}
Recall that a greedy solution to MDP would just maximize the immediate reward $\E{r_{\ell+1}(I_{\ell+1})}$ after $\ell$ downloads. 
\begin{thm}
For a given completely utilizing $\alpha$-$(V,R)$ storage scheme, the adaptive work-conserving scheduler that maximizes the expected immediate reward $\E{r_{\ell+1}(I_{\ell+1})\given I_\ell}$ after $\ell$ downloads is given by
\EQN{
\label{eqn:AlgoGreedy}
\Psi_{\Phi}(I_\ell)(b) = \arg\min_{v \in S_b^{\ell}} \rho_\ell^g(v), \quad \ell \in \{0, \cdots, V-1\},
}
where the greedy ranking function  $\rho_\ell^g(v)$ for a fragment $v$ after $\ell$ downloads is defined as 
\EQN{
\label{eqn:RankGreedy}
\rho_\ell^g(v) \triangleq \sum_{b \in \Phi_v} \SetIn{\abs{S_b^{\ell}}=1}.
}
\end{thm}
\begin{IEEEproof}
After each download $\ell$, 
maximizing the immediate reward of number of useful servers after next download $\E{N(I_{\ell+1})| I_\ell}$ 
is equivalent to minimizing the expected additional number of servers that become useless after next download i.e., $N(I_\ell)-\E{N(I_{\ell+1})| I_\ell}$. 
Conditioned on the set of downloaded fragments $I_\ell$, 
the number of useful servers $N(I_\ell)$ is deterministic. 
Therefore, we can write the conditional expectation of the reduction in the number of useful servers after $(\ell+1)$ downloads, as 
\EQ{
\E{N(I_\ell) - N(I_{\ell+1})\given I_\ell} 
= \sum_{b \in U(I_\ell)}\E{\SetIn{S_b^\ell = \set{v_{\ell+1}}}\given I_\ell},
}
where $v_{\ell+1}$ is the $(\ell+1)$th downloaded fragment. 
Note that $v_{\ell+1}$ is a random variable given $I_\ell$, 
and it takes value among all scheduled fragments with probability distribution given by Eq.~\eqref{eqn:ProbDownloadFragment}. 
We can re-write the conditional expectation of the reduction in number of useful servers given $\ell$ downloads, as 
\eq{
\E{N(I_\ell) - N(I_{\ell+1})\given I_\ell} 
&= \sum_{v \notin I_\ell}p_{I_\ell,I_\ell\cup\set{v}}\sum_{a \in \Phi_v}\SetIn{\abs{S_a^\ell} = 1}.
}
The above sum is a convex combination of greedy rank $\rho^g_{\ell}(v)$ over the probability distribution of scheduled fragments $v$. 
It follows that the greedy algorithm must schedule the fragment with the lowest greedy rank, 
at each useful server $b \in U(I_\ell)$ after $\ell$ downloads. 
\end{IEEEproof}
\begin{rem}
Note that the greedy rank $\rho_\ell^g(v)$ is equal to the number of servers that become useless if fragment $v$ gets downloaded at the $(\ell+1)$th download instant. 
\end{rem}

\subsubsection{Ranked schedulers}
Recall that a scheduling algorithm has to schedule a remaining fragment to be downloaded at each of the useful servers,  at each download instant. 
The greedy scheduler discussed previously, 
computes a function $\rho_\ell^g(v )$ at each download instant $\ell$ for each remaining fragment $v$ and schedules the fragment with smallest rank at each useful server.  
Instead of $\rho_{\ell}^g(v)$, we could consider other functions for remaining fragments giving us a class of algorithms for various rank functions.  
The rank function $\rho_{\ell}$ quantifies the suitability of the fragment to be scheduled for download. 
Fragments with lower rank are prioritized over fragments with higher rank while they are scheduled for downloading. 
The complete algorithm for a choice of $\rho_\ell: [V]\setminus I_\ell \to \R$ is given below. 

\begin{algorithm}
\caption{Suboptimal adaptive work-conserving scheduler }
\label{Algo:SuboptimalScheduler}
\begin{algorithmic}[1]
\REQUIRE $\alpha$-$(V,R)$ replication storage scheme $\Phi$ 
\ENSURE $((\Psi_\Phi(I_\ell)(b): b \in U(I_\ell): \ell \in \set{0, \dots, V-1})$
\STATE Set $S_b^{0} = S_b = \set{v \in [V]: b \in \Phi_v}$ for all $b$ in $[B]$
\FOR {$\ell \in \set{0, \cdots, {V-1}}$, } 
\FOR{$b$ in $U(I_\ell)$}
\STATE $\Psi_{\Phi}(I_\ell)(b) = \arg \min_{v \in S_b^\ell}\rho_\ell(v)$ 
\COMMENT Ties can be broken randomly or by any other rule.
\ENDFOR
\STATE Update $S_b^{\ell+1} = S_b^{\ell}\setminus\set{v_{\ell+1}}$ for all $b$ in $U(I_{\ell+1})$
\ENDFOR
\end{algorithmic}
\end{algorithm}

One big issue with the greedy approach is that it does not optimize the expected number of useful servers over all downloads. 
Since, it is oblivious to the evolution of $I_\ell$ and the choices it makes can steer the algorithm in a direction that does not minimize the mean download time. 
In particular, in the initial stages of download when $\ell$ is small, 
the greedy ranking function $\rho_{\ell}^g(v)=0$ for many fragments and the likelihood of not making the optimal choice is high. 

A better choice for the ranking 
function should be more sensitive to the download sequence and be able to assign a nonzero value even in the initial stages. 
This implies that a good ranking function must have some desirable properties. 
In the following discussion we attempt to derive some of them. 
First, we make the simplifying assumption that the ranking function $\rho_\ell(v)$ for a remaining fragment $v \notin I_\ell$ only depends on the collection 
$\{S_b^\ell: b \in \Phi_v \}$.

Recall that our goal is to maximize the number of useful servers, 
not just at the $\ell$th download  but over all the subsequent downloads. 
Intuitively, a choice of the metric $\rho(v)$ that ensures scheduling a fragment which is less likely to lead to servers storing single fragments in the future, might perform better compared to a greedy approach.

To this end, we make the following two observations. 
First observation is that a server becomes useless if all its stored fragments are downloaded. 
That is, $N(I_{\ell+1}) = \sum_{b\in [B]}\SetIn{S_b^{\ell+1} \neq \emptyset}$.
The second observation is that a fragment download reduces the number of remaining fragments on each of the server it is stored. 
If a remaining fragment $v \notin I_\ell$ is downloaded next, 
then 
\EQ{
\abs{S_b^{\ell+1}} = \abs{S_b^\ell} - \SetIn{b \in \Phi_v},\quad b \in U(I_\ell).
}
This suggests that we want to schedule fragments that are stored on servers with large cardinality of remaining fragment set. 
This implies the following monotonicity property for a good ranking function. 
\begin{defn}[Monotonicity] 
Consider two fragments $v,w \notin I_\ell$ after $\ell$ downloads, with the corresponding collection of remaining fragment sets $\set{S_b^\ell: b \in \Phi_v}$ and $\set{S_b^\ell: b \in \Phi_w}$ respectively. 
If there is a bijection $f: \Phi_w \to \Phi_v$ such that $\abs{S_b^\ell} \le \abs{S_{f(b)}^\ell}$ for all servers $b \in \Phi_w$, then $\rho_\ell(v) \le \rho_\ell(w)$. 
\end{defn}
\begin{rem}
We note that greedy ranking function $\rho^g_\ell$ has the monotonic property, however, it maps to zero for many fragments for small $\ell$. 
\end{rem}

Unfortunately, there are many collections of remaining fragment sets which can not be compared. 
We still need to rank such fragments, 
and therefore we propose the following harmonic ranking function $\rho_\ell^h: I_\ell^c \to \R$ at the $\ell$th download instant, 
for each of the remaining fragments $v \notin I_\ell$, 
as
\EQN{
\label{eqn:rankFunction}
\rho^h_\ell(v) \triangleq \sum_{a \in  \Phi_v} \frac{1}{\abs{S_a^\ell}},\quad v \notin I_\ell, \quad \ell \in \set{0, \dots, V-1}.
}
\begin{rem}
Note that the ranking function $\rho_\ell^h$ has the monotonic property, 
and the value $\rho_\ell^h(v)$ for a fragment $v$  is the harmonic sum of the cardinality of the set of remaining fragments $\abs{S_a^{\ell}}$ at each server $a \in \Phi_v$ after $\ell$ downloads. 
\end{rem}

Intuitively, one would like to schedule fragments that are present at servers with large remaining fragment sets. 
This ensures that scheduling of this fragment doesn't lead to large reduction in the number of useful servers, 
not only in the next download  instant but also in near future. 
Our proposed metric captures this effect in the following sense. 
That is, if the number of remaining fragments after $\ell$ downloads, at these servers are small, then we would not like to schedule this fragment. 

As the reciprocals of numbers increases steeply as the numbers get smaller, this harmonic sum is highly sensitive to low size of remaining fragment set at servers. 
That is, the rank of a fragment increases significantly if the servers on which it is hosted has very few remaining fragments to be downloaded.
Then, scheduling the fragment with the least value of the ranking function helps in decreasing the download probability of fragments which are stored on servers with low number of remaining fragments.  
This in turn reduces the probability of reduction of number of useful servers at each download instant. 

Algorithm~\ref{Algo:SuboptimalScheduler} is computationally efficient, easy to implement, 
and requires us to only keep track of the downloaded fragments for a given completely utilizing $\alpha$-$(V,R)$ storage scheme.
We know that ideally, the best algorithm should provide a solution that is jointly optimal over all download instants.
But, this algorithm only provides the best solution for the current download instant and does not take into account the impact of the current scheduling decision on the future evolution of the system.
Yet, we observe that its performance is comparable  with the optimal backward induction algorithm when implemented for a small number of fragments(see Fig.~{\ref{fig:MDP}, Appendix~\ref{app:BackwardInduction}}). 

We conclude this section with a few remarks concerning the complexity of Algorithm~\ref{Algo:SuboptimalScheduler}.
At every download instant, Algorithm~\ref{Algo:SuboptimalScheduler} computes the metric $\rho_\ell(v)$ for the remaining $V-\ell$ fragments to be downloaded. 
Assume that computation of $\rho_\ell(v)$ requires $c$ computations.
Since there are $V-\ell$ fragments, the cost of computing $\rho_\ell$ at each step is 
$c(V-\ell)$.
Finding the minimum at each server $b$ takes $|S_b^\ell|-1$ computations, giving 
$(V-\ell)R-N(I_{\ell})$ computations. 
Thus at each step $\ell$, the complexity is given by $O((c+R)(V-\ell) - N(I_\ell))$.
Summing over all $\ell$, we obtain an upper bound on the complexity of the algorithm as 
 $O(V^2(c+R))$.
 When the rank is given by $\rho_\ell^h(v)$, the cost $c=O(R)$.
 In this case, the complexity can be upper bounded by  $O(V^2R)$.

	\section{Randomized storage ensemble}
\label{sec:RandomCode}

We propose a random storage scheme for a $(VR,V)$ replication code stored over $B$ servers, 
where the storage capacity at each server is the entire $VR$ fragments. 
When the number of fragments $V$ grows large, 
we show that the fraction of fragments stored per server converges to $\alpha \triangleq \frac{R}{B}$ almost surely. 
Furthermore, we show that a \emph{typical}\footnote{A typical storage scheme refers to an element in a high probability set in the random ensemble.} storage scheme from the random ensemble achieves the universal upper bound on the expected number of useful servers aggregated over all fragment downloads, 
for any $\alpha$-$(V,R)$ replication storage scheme. 

Hence, a typical storage scheme from this random ensemble together with any work-conserving scheduling policy  solves the Problem~\ref{prob:MeanServiceCompletionTime} asymptotically as the number of fragments grows large. 
However, we observe that any storage scheme from the random replication storage code ensemble will not be a $\alpha$-$(V,R)$ replication code, 
with finite probability.

\subsection{Randomized replication coded storage}
A storage scheme for a $(VR, V)$ replication code, 
stored on a system with $B$ servers each having storage capacity of $VR$ fragments, 
is called  a \emph{$(B,V,R)$ replication storage scheme}. 
\begin{defn}%[Random replication storage scheme]
\label{defn:RandomPlaceRep} 
A $(B,V,R)$ replication storage scheme, 
where $r$th replica of fragment $v$ is stored on server $\Theta_{v,r} \in [B]$, chosen independently and uniformly at random from $B$ servers, 
is called a \emph{randomized $(B,V,R)$ replication storage scheme}. 
The collection of all random $(B,V,R)$ replication storage schemes is referred to as the \emph{random $(B,V,R)$ replication storage ensemble.} 
\end{defn}
\begin{rem}
A randomized $(B,V,R)$ replication storage scheme can be determined by \iid random vectors $\Theta_v = (\Theta_{v,r}: r \in [R]) \in [B]^R$ for all fragments $v \in [V]$,  
where 
\EQ{
P\set{\Theta_{v,r} = b} = \frac{1}{B},~\text{ for all } b \in [B].
}
We observe that since the number of replicated fragment $VR$ is smaller than the total system storage capacity $BVR$ for $B > 1$, 
this is a non completely utilizing storage scheme.
\end{rem}
Note that, the random vector $\Theta_v$ is not a set but a vector, 
since more than one replica of a file fragment can be stored on a single server. 
For each fragment $v$, we can compute the number of replicas of this fragment stored at server $b$ as
\EQN{
\label{eqn:}
\beta_{vb} \triangleq 
\sum_{r \in [R]}\SetIn{\Theta_{v,r} = b}.
}
\begin{lem}
For the random $(B,V,R)$ replication storage scheme defined in Definition~\ref{defn:RandomPlaceRep}, we have 
\EQ{
P(\cup_{ b \in [B]}\set{\beta_{vb} \ge 2}) \ge 1 - e^{\frac{-\alpha(R-1)}{2}},\quad v \in [V].
}
\end{lem}
\begin{IEEEproof}
The event of no server storing more than a single replica of a fragment $v$ in the random replication storage scheme is given by 
\begin{equation*}
E \triangleq \set{\Theta_{v,1} \neq \Theta_{v,2} \neq \dots \neq \Theta_{v,R}}.
\end{equation*}
Since placement of each replica $r \in [R]$ for each fragment $v \in [V]$ is \iid uniform over servers in $[B]$, 
the probability of this event is 
\begin{equation*}
P(E) 
=  \prod_{r=1}^{R-1}\Big(1 - \frac{r}{B}\Big)
\le e^{-\sum_{r=0}^{R-1}\frac{r}{B}}
= e^{-\frac{\alpha(R-1)}{2}}.
\end{equation*}
The result follows as the event $\cup_{ b \in [B]}\set{\beta_{vb} \ge 2}$ is the complement of the event $E$.
\end{IEEEproof}
As the choice of fragment $v$ in the above Lemma was arbitrary, 
it implies that the probability of each file fragment repeating on some server is finite. 
However, we can construct an occupancy set $\Phi_v$ for each fragment $v$ from this vector $\Theta_v$, 
by throwing away the repeated entries. 
That is, 
\EQ{
\Phi_v 
= \set{b \in [B]: \Theta_{v,r} = b\text{ for some }r \in [R]}.
}
This implies that $\abs{\Phi_v} \le R$, 
and this inequality is strict if any entry in the vector $\Theta_v$ is repeated twice. 

We will consider a family of $(B,V,R)$ random replication storage schemes for increasing values of number of fragments $V$, while keeping the ratio $\alpha = \frac{R}{B}$ constant for the system. 
In this case, 
we will show that the fraction of file fragments stored by each server converges to $\alpha$ for the proposed random $(B,V,R)$ replication storage scheme.  
Recall that the ratio $\alpha$ is the normalized storage capacity of each servers in a completely utilizing $\alpha$-$(V,R)$ storage scheme. 

The normalized number of fragments stored at any server $b \in [B]$ is defined as 
\EQN{
\label{eqn:NumFragmentsRandom}
\alpha_b^{\rep} \triangleq \frac{1}{V}\sum_{v \in [V]}\sum_{r \in [R]} \SetIn{\Theta_{v,r} = b}.
}
\begin{rem} 
If the normalized number of fragments $\alpha_b^\rep = \alpha$, then the randomized $(B,V,R)$ replication storage scheme defined in Definition~\ref{defn:RandomPlaceRep} in terms of \iid vectors $(\Theta_v: v \in [V])$\footnote{
Even though $\Theta: [V] \to [B]^R$ is a collection of \iid random variables, 
we observe that normalized number of pieces on each server $\alpha_b^\rep$ are dependent random variables. 
To see this, we observe that 
$P\set{\Theta_{v,r}=b, \Theta_{v,r} = a} = 0 \neq \frac{1}{B^2}$ for any $b \neq a$. 
It follows that 
\EQ{
\E{\alpha_b^\rep \alpha_a^\rep} = \alpha^2 - \frac{1}{B}\alpha \neq \alpha^2 = \E{\alpha_b^\rep}\E{\alpha_a^\rep}.
}
}, is an underutilizing $\alpha$-$(V,R)$ storage scheme, 
since there exists servers storing redundant replicas of the same fragment with high probability. 
\end{rem} 

\begin{defn}
\label{defn:AsymptoticBVRKStorage}
We say that a randomized $(B, V, R)$ replication storage scheme is an $\alpha$-$(V, R)$  storage scheme asymptotically in $V$, if 
for each server $b$, $\lim_{V \to \infty}\alpha_b^\rep = \alpha$ almost surely.
\end{defn}

\begin{thm}
\label{thm:replication_serverstorage} 
 The randomized $(B, V, R)$ storage scheme defined in Definition~\ref{defn:RandomPlaceRep}  
is an $\alpha$-$(V, R)$ storage scheme  asymptotically in $V$. 
\end{thm}
\begin{IEEEproof} 
We first a define a random sequence $(M_n : n\le V)$, 
such that the $n$th term is 
\EQ{
M_n \triangleq \frac{1}{V}\sum_{v =1}^{n}\sum_{r \in [R]} \Big(\SetIn{\Theta_{v,r} = b} - \E{\SetIn{\Theta_{v,r} = b}}\Big),~n \le V.
}

Since each term in the above summation is \iid and zero mean, 
it can be easily verified that the sequence $(M_n: n\le V) $ is a Martingale. 
Further, 
we can write the second moment of $M_n$ for $n \le V$ as 
\EQ{
\E{M_n^2} = \frac{nR}{V^2}\Var\Big(\SetIn{\Theta_{v,r} = b}\Big) 
=  \frac{nR}{V^2B}\Big(1 - \frac{1}{B}\Big) \le \frac{\alpha}{V}.
}
This implies that as number of fragments $V$ grows to infinity keeping the fraction $\alpha = {R}/{B}$ constant,   
the Martingale $(M_n: n \le V)$ converges to zero in mean square sense.
Furthermore, as the variance of a random variable is always non-negative, we see that $\E{\abs{M_n}} \le (\E{M_n^2})^{1/2}$.
Thus, 
\EQ{
0 \le \limsup_{V\to \infty }\sup_{n \le V}\E{\abs{M_n}} \le \limsup_{V\to\infty}\sqrt{\frac{\alpha}{V}} = 0.
}
In particular, it implies that $\lim_{V\to\infty}\E{\abs{M_V}} = 0$, 
or equivalently $\lim_{V\to\infty}M_V = 0$ almost surely. 
The result follows since $M_V = \alpha_b^\rep - \alpha$. 
\end{IEEEproof}

We can compute the sum of mean number of useful servers aggregated over all downloads, 
when the mean is taken over ensemble of random $(B, V, R)$ replication storage schemes.
\begin{thm}
\label{thm:performance_random_rep}
For the random  $(B,V,R)$ replication storage ensemble  defined in Definition~\ref{defn:RandomPlaceRep}, 
we can write
\EQ{
\frac{1}{BV}\sum_{\ell = 0}^{V-1} \E{N(I_{\ell})} = 1- \frac{\Big(1 - \frac{1}{B}\Big)\Big(1 - (1-\frac{1}{B})^{RV}\Big)}{V\Big(1 - (1-\frac{1}{B})^R\Big)}.
}
\end{thm}
\begin{IEEEproof}
For any storage scheme in random $(B, V, R)$ replication storage ensemble, the storage is specified by 
the random vectors $\Theta = (\Theta_v: v \in [V])$. 
Since $(\Theta_{v,r}: r \in [R])$ is an \iid sequence, 
we can write the ensemble probability that none of the replicas of a fragment $v$ are stored on server $b$ as 
\EQ{
P(\cap_{r \in [R]}\set{\Theta_{v,r} \neq b}) = \left(1- \frac{1}{B}\right)^R.
}
A server $b \in U(I_\ell)$ is useful after download of $\ell$-fragment subset $I_\ell$, if and only if $b = \Theta_{v,r}$ for some $v \notin I_\ell$ and $r \in [R]$. 
Since each fragment was stored independently and uniformly at random, 
we obtain the ensemble probability of a server $b$ not being useful,  
given download sequence $I_\ell$, as 
\EQ{
P\set{b \notin U(I_\ell)} 
= P(\bigcap_{v \notin I_\ell}\bigcap_{r \in [R]}\set{\Theta_{v,r}\neq b})
= \Big(1 - \frac{1}{B}\Big)^{R(V - \ell)}.
}
Since $N(I_\ell) = \sum_{b \in [B]}\SetIn{b \in U(I_\ell)}$, 
it follows from the linearity of expectation that 
\EQN{
\label{eqn:NumUsefulServersRandom}
\frac{1}{BV}\E{N(I_{\ell})} = \frac{1}{V}\Big[1 - \Big(1 - \frac{1}{B}\Big)^{R(V - \ell)}\Big].
}
Result follows from summing the above equation on both sides over  $\ell \in \set{0, \dots, V-1}$. 
\end{IEEEproof}
\begin{rem} 
Recall that we did not make any assumption on scheduling policies for the above computation, 
other than that it is a work-conserving scheduling policy. 
We assumed that a scheduler will only schedule a remaining fragment. 
The ensemble averaging allowed us to get rid of combinatorial constraints on work-conserving scheduling posed by the storage schemes. 
\end{rem}
\begin{cor}
\label{cor:RandomRepAsympOptimal}
A random $(B,V,R)$ replication storage scheme is  asymptotically optimal solution to the  Problem~\ref{prob:MeanServiceCompletionTime} almost surely. 
\end{cor}
\begin{IEEEproof} 
As the number of fragments $V$ tends to infinity, 
the following two results hold. 
\begin{compactenum}
\item By Theorem~\ref{thm:replication_serverstorage},
 a proposed random $(B,V,R)$
replication storage scheme almost surely converges to a underutilizing $\alpha$-$(V,R)$ replication storage scheme. 
\item From Theorem~\ref{thm:performance_random_rep}, we obtain
\EQ{
\lim_{V\to\infty}\frac{1}{BV}\sum_{\ell=0}^{V-1}\E{N(I_\ell)} = 1.
}
\end{compactenum}
From Eq.~\eqref{eqn:NormalizedSumUsefulServersUB} and Remark~\ref{rem:underutilizingUB}, we observe that the ensemble mean of normalized number of useful servers for the 
proposed random $(B,V,R)$ replication storage scheme meets the upper bound for any $\alpha$-$(V,R)$ replication storage scheme, 
asymptotically in $V$.
Thus, we see that the proposed random storage scheme asymptotically achieves the universal upper bound almost surely, and is the solution to Problem~\ref{prob:MeanServiceCompletionTime}. 
\nix{
Taking expectation on both sides of Eq.~\eqref{eqn:NormalizedSumUsefulServersUB}, 
we know that the normalized upper bound on the expected number of useful servers aggregated over all downloads,
for any completely utilizing $\alpha$-$(V,R)$ replication storage scheme is unity. 
}
\end{IEEEproof}

\section{Comparison with MDS codes}
\label{sec:ComparisonMDS}
So far we have looked at storage schemes based on $(VR, V)$ replication codes.
In this section we consider storage schemes based on $(VR,V)$ MDS codes assuming that the field is large enough so that a $(VR, V)$ code exists.
MDS codes are known to outperform replication codes in many settings. 
For example, MDS codes have better code rates for same fraction of erasure correction~\cite{Peterson1972}, 
and are shown to be latency optimal for class of symmetric codes in single fragment storage~\cite{Badita2019TIT}.

In this section, we first show that 
among all $\alpha$-$(V, R)$ coded storage schemes, the ones based on MDS codes 
minimize the mean download time. 
Second, we find the bounds on the number of useful servers for MDS coded storage, 
and show that replication coded storage is asymptotically order optimal. 
That is, when the number of fragments grows large, 
the average number of useful servers per fragment can be achieved by random replication. 
Third, we show that when each server can store whole file, i.e. $K \ge V$, 
then the replication coded storage is as good as a MDS coded storage, 
even in non-asymptotic regime. 
\begin{defn}
Consider a file with $V$ fragments encoded to $VR$ coded fragments, 
and completely utilizing storage of this $(VR, V)$ code on a $\alpha$-$B$ system. 
Such storage schemes are referred to as \emph{$\alpha$-$(V,R)$ coded storage schemes}, 
where the normalized storage capacity per server is $\alpha = R/B = K/V$ and the code rate is $1/R$. 
\end{defn}
\begin{rem}
\nix{Consider a file with $V$ fragments, that can be encoded into a $VR$ coded fragments, and stored on a storage system with $B$ servers that can store $K$ coded fragments each. 
That is, we are considering a $(VR, V)$ code stored on a $(B,V,R,K)$ system, }
For any $\alpha$-$(V,R)$ coded storage scheme, 
the number of useful servers $N(I_\ell)$ after $\ell$ downloads is always upper bounded by the total number of servers $B$, and hence 
\EQN{
\label{eqn:AvgUsefulServersUbMds}
\frac{1}{BV}\sum_{\ell=0}^{V-1}N(I_\ell) \le 1.
}
\end{rem}
\begin{defn}
Any $V$ subset of $VR$ coded fragments that suffices to 
decode a $(VR,V)$ code, i.e.,  reconstruct the $V$ uncoded fragments,  is called an \emph{information set}\cite{Gopalan2016SODA,Badita2019TIT}.
For an $\alpha$-$(V,R)$ coded storage scheme, 
we can define the collection of all information sets~\cite[Section II]{Badita2019TIT}, as $\cI$.  
\end{defn}
For a completely utilizing $\alpha$-$(V,R)$ replication storage scheme,  
information sets consist of distinct $V$ fragments. 
For a completely utilizing $\alpha$-$(VR, V)$ MDS coded storage, information sets are any $V$ coded fragments, 
and hence $\cI = \set{S \subset [VR]: \abs{S} = V}$. 
This implies that the collection of information sets for MDS code includes collection of information sets for any other $(VR, V)$ code.

\begin{thm}
Among all $\alpha$-$(V,R)$ coded storage schemes, 
MDS codes minimize the mean download time. 
\end{thm}
\begin{IEEEproof}
For any completely utilizing $\alpha$-$(V, R)$ coded storage scheme, 
each server $b$ is storing  a set $S_b \subseteq [VR]$ of $\abs{S_b} = \alpha V$ out of $VR$ coded fragments. 
Further, the first $\ell$ downloaded symbols $I_\ell$ are an $\ell$-subset of some information set $S \in \cI$. 
Then, we can write the set of useful servers after $\ell$ downloads as those that have the remaining coded symbols for such information sets. 
That is, 
\EQ{
U(I_\ell) = \cup_{S \in \cI}\set{b \in [B]: (S\setminus I_\ell) \cap S_b \neq \emptyset}.
}
Recall that the collection of information sets for MDS codes includes collection of information sets for any other $(VR,V)$ code. 
Using this fact together with the definition of the set of useful servers, 
it follows that the largest possible set of useful servers among all $\alpha$-$(V,R)$ coded storage schemes,  
is the one achieved by MDS coded storage for the same download sequence $I_\ell$. 
From coupling arguments for the download sequence and Eq.~\eqref{eq:MeanDownloadTime} for mean download time, 
the result follows by induction on the number of downloaded fragments. 
\end{IEEEproof}

\subsection{Asymptotic order optimality of replication codes}
We established that the among all $\alpha$-$(V,R)$ coded storage schemes, 
an MDS code  has the largest number of useful servers.  
we next find bounds on the number of useful servers for MDS coded storage, which can be used as a benchmark to compare replication coded storage. 
\begin{lem}
For a completely utilizing $\alpha$-$(V,R)$ MDS storage scheme, the number of useful servers $N^{\mds}(I_{\ell})$ is bounded as 
\EQ{
B - \floor{\frac{\ell}{K}} \le N^\mds(I_\ell) \le \min(B, VR - \ell).
}
\end{lem}
\begin{IEEEproof}
For an $\alpha$-$(V,R)$ MDS coded storage scheme, each $VR$ coded fragment is useful and downloading any $V$ coded fragments suffices to reconstruct the entire file.
Therefore, if $\ell < V$ fragments are downloaded, 
then all the servers that store any of the remaining $VR - \ell$ fragments are useful.  
As each server can store $K=\alpha V$ fragments each, 
a server can become useless if and only if all its $K$ coded fragments have been downloaded. 
Therefore, the maximum number of servers that can be useless after $\ell$ downloads is $\floor{\ell/K}$. 

To obtain the upper bound on the number of useful servers, we make the following observations. 
First, that all servers remain useful at the $\ell$th download if less then $K$ coded fragments have been downloaded from them. 
However, if the number of remaining coded fragments $VR - \ell < B$, 
then at most $VR-\ell$ unique servers are useful. 
\end{IEEEproof}

The previous lemma will immediately give us the following result by taking average over 
all $V$ fragments. 

\begin{cor}
For a completely utilizing $\alpha$-$(V,R)$ MDS coded storage scheme 
with code rate $\frac{1}{R} \le \frac{V}{B+V}$, the normalized aggregate number of useful servers is bounded as 
\EQN{
\label{eqn:AvgUsefulServersUbMds}
1 - \frac{1}{2R}(1 - \frac{1}{V})\le \frac{1}{BV}\sum_{\ell=0}^{V-1} N^\mds(I_\ell) \le 1.
}
\end{cor}

\begin{rem}
We recall from Theorem~\ref{thm:replication_serverstorage} that a random $(B,V,R)$ replication storage scheme achieves a storage fraction $\alpha = R/B$ on each server, as the number of fragments $V$ becomes large. 
Further, from Corollary~\ref{cor:RandomRepAsympOptimal} we observe that the limit of average number of useful servers for a typical random $(B,V,R)$ replication storage scheme meets the upper bound for MDS codes in Eq.~\eqref{eqn:AvgUsefulServersUbMds}, as the number of fragment grows. 
This implies that replication coded storage is asymptotically order optimal. 
\end{rem}

\subsection{Optimality of replication codes for large storage} 
\label{subsec:LargeStorage} 

So far, we have considered the case $\alpha \le 1$. 
In other words, it means that storage capacity per server is $K = \alpha V \le V$, i.e. each server can store at most a single file. 
We now show that, when $K \ge V$, 
then there exists a $(VR, V)$ replication code such that the number of useful servers remains $B$ after every download, for any work-conserving scheduling policy.  
Thus, the average number of useful servers for replication code meets the universal upper bound for all fragment size $V$, 
when the storage can store at least the entire file.  

\begin{lem}
\label{lm:repStorageOtpimality}
There exists a $(VR, V)$ replication code,  
that meets the universal upper bound on average number of useful servers for any completely utilize $\alpha$-$(V,R)$ coded storage scheme when $\alpha \ge 1$.   
\end{lem} 
\begin{IEEEproof} 
Since $VR = BK$ and $K \ge V$, it follows that $R = B\frac{K}{V} \ge B$. 
Thus, we can construct replication mappings $\theta_v: [R] \to [B]$ for each fragment $v \in [V]$, 
where $\theta_{vr}$ is the server on which $r$th replica of fragment $v$ is stored. 
Let $\theta_{v r} = r \in [B]$ for all replica $r \le B$ and fragment $v \in [V]$. 

There are $(R-B)V$ replicas left to be stored on $B(K-V)$ places on $B$ servers. 
 As $BK = VR$, we get $(R-B)V = K(K-V)$ and can choose to store the remaining replicas at these remaining locations in any order.

It follows that each of $V$ fragments is stored on all the $B$ servers, 
and hence all $B$ servers remain useful until all fragments are downloaded. 
\end{IEEEproof}

	\section{Numerical studies}
\label{sec:sims}
    In this section, we present the results of our numerical studies for 
    completely utilizing $\alpha$-$(V, R)$ replication
    storage schemes and work-conserving scheduling policies.
    The interplay of the storage scheme and work-conserving scheduling policies determines the overall 
     download time for a file. 
    We use the storage schemes proposed in Section~\ref{sec:DesignBasedStorage}.
    We study the performance of these storage codes in conjunction with various nonadaptive and adaptive scheduling policies. 
    
    Before presenting the numerical results, we first review some  scheduling policies
    and  illustrate them  by considering a completely utilizing $\frac{3}{7}$-$(7, 3)$ replication storage scheme.
    We consider the storage scheme constructed from a projective plane. 
    
    \begin{compactenum}
    
    \item {\em Smallest index first scheduling}: 
        A straightforward nonadaptive scheduling policy is to schedule the fragments based on their indices. 
    We could arrange the fragments in the increasing order of the fragment index. 
    After a fragment is downloaded, the fragment with the next highest index is moved to the head of the server. 
    This scheduling policy applied to the storage scheme in Table~\ref{tab:design-7733}
    is shown in Table~\ref{tab:design-ascending}.
    
    \begin{table}[htbp]	
	\begin{center}
		\begin{tabular}{c|c|c|c|c|c|c|c}
			\hline \hline
			\multicolumn{8}{c}{Smallest index first scheduling}\\
			\hline
			Server   & 1 & 2 & 3 & 4 & 5 & 6 & 7 \\ \hline
			Layer 1  & 1 & 3 & 1 & 1 & 2 & 3 & 2 \\ 
			Layer 2  & 2 & 4 & 5 & 4 & 5 & 6 & 4 \\ 
			Layer 3  & 3 & 5 & 6 & 7 & 7 & 7 & 6 \\ \hline
			\hline
		\end{tabular}
	\end{center}
	\caption{A static scheduling policy for the $\frac{3}{7}$-$(7, 3)$ replication storage coding scheme with the fragments scheduled in increasing order of their indices.}
	\label{tab:design-ascending}
\end{table}

    \item {\em Uniform diversity scheduling}: The previous scheduling policy,   leads to 
    an asymmetric scheduling in that all fragments are not equally distributed at the heads of the servers. 
    For instance, in Table~\ref{tab:design-ascending}, fragments $4$ to $7$ are not scheduled in the first layer. 
    We can make the policy more symmetric, by scheduling as many distinct fragments at the head of each server and in each subsequent layer. 
    The motivation being that having a diversity of fragments at the heads of the servers leads to larger number of fragments being downloaded in parallel. 
    One such scheduling for the design based $\frac{3}{7}$-$(7, 3)$ replication storage is shown in Table~\ref{tab:design-matching}.
    For storage schemes based on projective planes and the cyclic shift storage scheme, where 
    $B=V$, it is always possible to place fragments in each layer across the servers as a permutation of all the fragments.
    Such a uniform scheduling may not  be possible for every storage scheme. 

\begin{table}[hhhh]
	\begin{center}
		\begin{tabular}{c|c|c|c|c|c|c|c}
			\hline \hline
			\multicolumn{8}{c}{Uniformly diverse scheduling}\\
			\hline
			Server   & 1 & 2 & 3 & 4 & 5 & 6 & 7 \\ \hline
			Layer 1  & 1 & 3 & 5 & 7 & 2 & 6 & 4 \\ 
			Layer 2  & 3 & 4 & 6 & 1 & 5 & 7 & 2 \\ 
			Layer 3  & 2 & 5 & 1 & 4 & 7 & 3 & 6 \\ \hline
			\hline
		\end{tabular}
	\end{center}
	\caption{A scheduling for $\frac{3}{7}$-$(7, 3)$  replication storage coding scheme with uniform diversity at each layer. }
	\label{tab:design-matching}
\end{table}
 
\item {\em Pushback scheduling}: 
This scheduling policy aims to maximize the number of useful servers toward the end when a large number of pieces have been downloaded. 

A heuristic scheduling policy that aims to maximize $N(I_\ell)$ in this range is as follows:
Pick a server $b$ and schedule the fragments of
$S_b$  last in the other servers i.e. $[B]\setminus \{b\}$. 

In a projective plane based storage scheme, the pushback policy will schedule the fragments stored in $b$th server on  $R(R-1)$ disjoint servers. 
Comparing with the 
bound given in Theorem~\ref{thm:LowerBoundNumUsefulUpperGen} for the number of useful servers,
we can see this will lead to larger number of useful servers at the end.

\begin{table}[hhhh]
	\begin{center}
		\begin{tabular}{c|c|c|c|c|c|c|c}
			\hline \hline
			\multicolumn{8}{c}{Pushback  with uniform diversity scheduling}\\
			\hline
			Server   & 1 & 2 & 3 & 4 & 5 & 6 & 7 \\ \hline
			Layer 1  & \bl{1} & 4 & 5 & 7 & 5 & 6 & 4 \\ 
			Layer 2  & \bl{3} & 5 & 6 & 4 & 7 & 7 & 6 \\ 
			Layer 3  & \bl{2} & \bl{3} & \bl{1} & \bl{1} & \bl{2} & \bl{3} & \bl{2} \\ \hline
			\hline
		\end{tabular}
	\end{center}
	\caption{Combining pushback policy with the uniform diversity policy for the $\frac{3}{7}$-$(7, 3)$ replication storage coding scheme. The fragments  of the first server (in blue) are placed in the last layer, due to which they are scheduled last in those servers.}
	\label{tab:matchings-onePushback}
\end{table}  

This policy can be combined with  any other scheduling policy. 
When combined with the  scheduling policy in Table~\ref{tab:design-matching}, the effective scheduling is as in Table~\ref{tab:matchings-onePushback} and 
 combining with the smallest index first policy in Table~\ref{tab:design-ascending} gives the  placement in Table~\ref{tab:ascending-onePushback}.

\begin{table}[htbp]
	\begin{center}
		\begin{tabular}{c|c|c|c|c|c|c|c}
			\hline \hline
			\multicolumn{8}{c}{Pushback with smallest index first scheduling}\\
			\hline
			Server   & 1 & 2 & 3 & 4 & 5 & 6 & 7 \\ \hline
			Layer 1  & \bl{1} & 4 & 5 & 4 & 5 & 6 & 4 \\ 
			Layer 2  & \bl{2} & 5 & 6 & 7 & 7 & 7 & 6 \\ 
			Layer 3  & \bl{3} & \bl{3} & \bl{1} & \bl{1} & \bl{2} & \bl{3} & \bl{2} \\ \hline
			\hline
		\end{tabular}
	\end{center}
	\caption{Combining pushback policy with the smallest index first policy in  Table~\ref{tab:design-ascending} for the $\frac{3}{7}$-$(7, 3)$ replication storage scheme.}
	\label{tab:ascending-onePushback}
\end{table}

\end{compactenum}

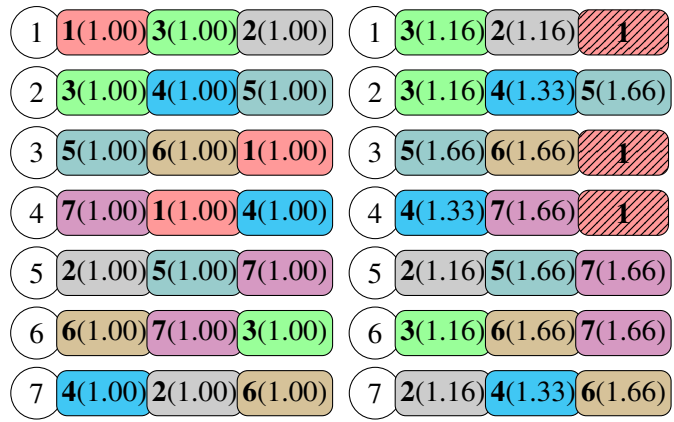
\begin{figure}[hhh]
\centerline{\scalebox{1}{\begin{tikzpicture}
[node distance=2cm, draw=black, thin, >=stealth', 
  req/.style={rectangle, rounded corners, minimum size=15mm, draw=black},
  vault/.style={circle, minimum size = 2cm, draw=black},
  server/.style={circle, minimum size = 0.5cm, draw=black},
  content/.style={rectangle, rounded corners, minimum height=1cm, minimum width=4cm, draw=black},
  cache/.style={rectangle, rounded corners, minimum height=0.6cm, minimum width=1.2cm, draw=black, inner sep=2pt},
  note/.style={rectangle, rounded corners, minimum size=5mm, inner sep=5pt, draw=black},
  DownloadedCache/.style={rectangle, rounded corners, minimum height=0.6cm, minimum width=1.2cm, draw=black, inner sep=2pt},
  ]											
%----------------------First Stage---------------------------
  %\node[note, fill=yellow!20] at (2.2,5.8) {First Stage};
 % \node[note, fill=yellow!20] at (-2,5.8) {Second Stage};

  \node[server](server1) at (0.1,4.8){1};
  \node[cache, fill=red!40] (cache11) at (1, 4.8) {\textbf{1}(1.00)};
  \node[cache, fill=green!40] (cache12) at (2.2, 4.8) {\textbf{3}(1.00)};
  \node[cache, fill=black!20] (cache13) at (3.4, 4.8) {\textbf{2}(1.00)};

  \node[server](server2) at (0.1,4){2};
  \node[cache, fill=green!40] (cache21) at (1, 4) {\textbf{3}(1.00)};
  \node[cache, fill=cyan!60] (cache22) at (2.2, 4) {\textbf{4}(1.00)};
  \node[cache, fill=blue!50!green!40] (cache23) at (3.4, 4) {\textbf{5}(1.00)};

  \node[server](server3) at (0.1,3.2){3};
  \node[cache, fill=blue!50!green!40] (cache31) at (1, 3.2) {\textbf{5}(1.00)};
  \node[cache, fill=green!40!red!40] (cache32) at (2.2, 3.2) {\textbf{6}(1.00)};
  \node[cache, fill=red!40] (cache33) at (3.4, 3.2) {\textbf{1}(1.00)};

  \node[server](server4) at (0.1,2.4){4};
  \node[cache, fill=blue!40!red!40] (cache41) at (1, 2.4) {\textbf{7}(1.00)};
  \node[cache, fill=red!40] (cache42) at (2.2, 2.4) {\textbf{1}(1.00)};
  \node[cache, fill=cyan!60] (cache43) at (3.4, 2.4) {\textbf{4}(1.00)};

  \node[server](server5) at (0.1,1.6){5};
  \node[cache, fill=black!20] (cache51) at (1, 1.6) {\textbf{2}(1.00)};
  \node[cache, fill=blue!50!green!40] (cache52) at (2.2, 1.6) {\textbf{5}(1.00)};
  \node[cache, fill=blue!40!red!40] (cache53) at (3.4,1.6) {\textbf{7}(1.00)};
  
  \node[server](server6) at (0.1,0.8){6};
  \node[cache, fill=green!40!red!40] (cache61) at (1, 0.8) {\textbf{6}(1.00)};
  \node[cache, fill=blue!40!red!40] (cache62) at (2.2, 0.8) {\textbf{7}(1.00)};
  \node[cache, fill=green!40] (cache63) at (3.4, 0.8) {\textbf{3}(1.00)};
  
  \node[server](server7) at (0.1,0){7};
  \node[cache, fill=cyan!60] (cache71) at (1, 0) {\textbf{4}(1.00)};
  \node[cache, fill=black!20] (cache72) at (2.2, 0) {\textbf{2}(1.00)};
  \node[cache, fill=green!40!red!40] (cache73) at (3.4, 0) {\textbf{6}(1.00)};

%----------------------Second Stage---------------------------
  \node[server](server1) at (4.6,4.8){1};

  \node[cache, fill=green!40] (cache12) at (5.5, 4.8) {\textbf{3}(1.16)};
  \node[cache, fill=black!20] (cache13) at (6.7, 4.8) {\textbf{2}(1.16)};
  \node[DownloadedCache, fill=red!40, postaction={pattern=north east lines}] (cache11) at (7.9, 4.8) {\textbf{1}};
  
  \node[server](server2) at (4.6,4){2};
  \node[cache, fill=green!40] (cache21) at (5.5, 4) {\textbf{3}(1.16)};
  \node[cache, fill=cyan!60] (cache22) at (6.7, 4) {\textbf{4}(1.33)};
  \node[cache, fill=blue!50!green!40] (cache23) at (7.9, 4) {\textbf{5}(1.66)};
  
  \node[server](server3) at (4.6,3.2){3};
  \node[cache, fill=blue!50!green!40] (cache31) at (5.5, 3.2) {\textbf{5}(1.66)};
  \node[cache, fill=green!40!red!40] (cache32) at (6.7, 3.2) {\textbf{6}(1.66)};
  \node[DownloadedCache, fill=red!40, postaction={pattern=north east lines}] (cache33) at (7.9, 3.2) {\textbf{1}};
  
  \node[server](server4) at (4.6,2.4){4};
  \node[cache, fill=cyan!60] (cache43) at (5.5, 2.4) {\textbf{4}(1.33)};
  \node[cache, fill=blue!40!red!40] (cache41) at (6.7, 2.4) {\textbf{7}(1.66)};
  \node[DownloadedCache, fill=red!40, postaction={pattern=north east lines}] (cache42) at (7.9, 2.4) {\textbf{1}};
  
  \node[server](server5) at (4.6,1.6){5};
  \node[cache, fill=black!20] (cache51) at (5.5, 1.6) {\textbf{2}(1.16)};
  \node[cache, fill=blue!50!green!40] (cache52) at (6.7, 1.6) {\textbf{5}(1.66)};
  \node[cache, fill=blue!40!red!40] (cache53) at (7.9,1.6) {\textbf{7}(1.66)};
  
  \node[server](server6) at (4.6,0.8){6};
  \node[cache, fill=green!40] (cache63) at (5.5, 0.8) {\textbf{3}(1.16)};
  \node[cache, fill=green!40!red!40] (cache61) at (6.7, 0.8) {\textbf{6}(1.66)};
  \node[cache, fill=blue!40!red!40] (cache62) at (7.9, 0.8) {\textbf{7}(1.66)};
  
  \node[server](server7) at (4.6,0){7};
  \node[cache, fill=black!20] (cache72) at (5.5, 0) {\textbf{2}(1.16)};
  \node[cache, fill=cyan!60] (cache71) at (6.7, 0) {\textbf{4}(1.33)};
  \node[cache, fill=green!40!red!40] (cache73) at (7.9, 0) {\textbf{6}(1.66)};

  \end{tikzpicture}}}
\caption{
We show first two steps of a sample path of download sequence for a design based $\frac{3}{7}$-$(7,3)$ replication storage scheme given in Table~\ref{tab:design-matching}. 
The ranked adaptive scheduling defined in Algorithm~\ref{Algo:SuboptimalScheduler} with harmonic rank function defined in Eq.~\eqref{eqn:rankFunction} is used. 
We have listed the rank of the fragment along with its identity.} 
\label{Fig:OneSamplePath}
\end{figure}

We now illustrate an example of an adaptive ranked scheduler with harmonic ranking function defined in Eq.~\eqref{eqn:rankFunction}, 
with design based $\frac{3}{7}$-$(7,3)$ replication storage code given in Table~\ref{tab:design-matching}. 
We will look at one sample path of download sequence in Fig.~\ref{Fig:OneSamplePath}. 
Initially harmonic rank of all fragments is identically unity, 
and any stored fragment can be scheduled at any of the $7$ servers. 
Let us assume that the first downloaded fragment is $1$. 
We now compute the harmonic rank of the remaining six fragments $\set{2,3,4,5,6,7}$ as $\set{1.16, 1.16, 1.33, 1.66, 1.66, 1.66}$ respectively. 
Thus, Algorithm~\ref{Algo:SuboptimalScheduler} schedules either fragment~2 or~3 on server~1,
fragment~3 on server~2, 
either fragment~5 or~6 on server~3, 
fragment~4 on server~4,
fragment~2 on server~5, 
fragment~3 on server~6, 
and fragment~2 on server~7.

We now present the results of our numerical studies on a completely utilizing $\alpha$-$(V, R)$ replication storage scheme constructed from a projective plane
of order $q=11$. 
This results in a symmetric $(VR,V)$ replication code stored on an $\alpha$-$B$ system, 
where the number of servers $B=V = q^2+q+1$, 
the replication factor of each fragment $R = q+1$, 
and the storage capacity of each server is a fraction $\alpha = R/B$ of all $V$ fragments. 
We also considered an alternative completely utilizing $\alpha$-$(V,R)$ replication storage scheme based on cyclic shift of fragments, 
for the identical parameters $\alpha, V, R$.  
As mentioned earlier, the download time for each fragment is modelled as an independent random variable that has an exponential distribution with rate $\mu = \num{e-5}$, 
chosen 
to amplify the differences between various storage schemes and scheduling policies. 
We performed Monte Carlo simulations of our system setup with $\num{1e5}$ runs.  
We computed the normalized empirical mean  $\hat{\mathbb{E}}[N(I_\ell)]/B$, of number of useful servers $N(I_\ell)$ after $\ell$ downloads, 
averaged over all simulation runs. 
Performance of various storage schemes and scheduling policies are compared by plotting the normalized empirical average $\hat{\mathbb{E}}[N(I_\ell)]/B$ as the fraction of downloads $\ell/V$ grows. 
From Eq.~\eqref{eq:MeanDownloadTime}, we know that the performance of the scheme is better if the number of useful servers remains high as the download progresses. 
That is, a uniformly higher plot is indicative of a better performance.

\begin{figure}[tbh]
\centerline{\scalebox{0.9}{\input{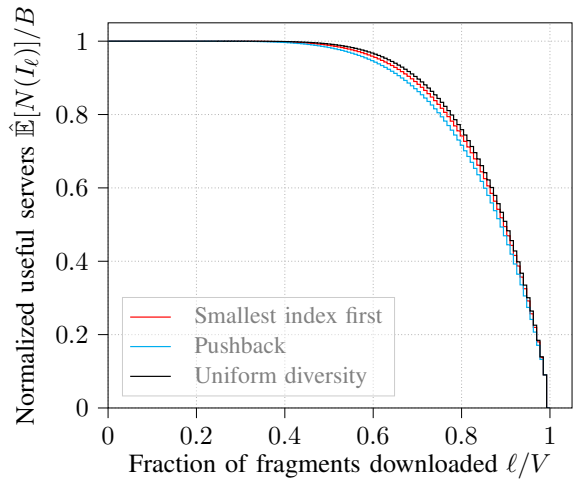}}}
\caption{
This plot shows an empirical average of the normalized number of useful servers $\hat{\mathbb{E}}[N(I_\ell)]/B$ for the cyclic shift based $\frac{12}{133}$-$(133,12)$ replication storage scheme with nonadaptive scheduling policies.
}
\label{fig:AvgUsefulServerslistCyclicSchemesExp}
\end{figure}

\begin{figure}[tbh]
\centerline{\scalebox{0.9}{\input{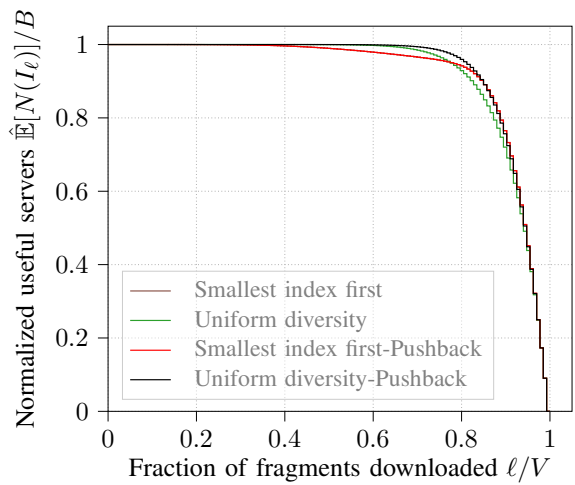}}}
\caption{
This plot shows an empirical average of the normalized number of useful servers $\hat{\mathbb{E}}[N(I_\ell)]/B$ for the projective plane based $\frac{12}{133}$-$(133,12)$ replication storage scheme with nonadaptive scheduling policies.
}
\label{fig:AvgUsefulServerslistStaticSchemesExp}
\end{figure}

The simulation results for the cyclic shift based storage scheme for nonadaptive scheduling policies, are shown in Fig.~\ref{fig:AvgUsefulServerslistCyclicSchemesExp}. 
We observe that the uniform diversity scheduling has the best performance among the proposed nonadaptive scheduling policies. 
The results for the projective plane based scheme for nonadaptive scheduling policies, 
are shown in Fig.~\ref{fig:AvgUsefulServerslistStaticSchemesExp}. 
In this case, we observe that the scheduling policy that combines the uniform diversity with pushback has the best performance. 

\begin{figure}[tbh]
\centerline{\scalebox{0.9}{\input{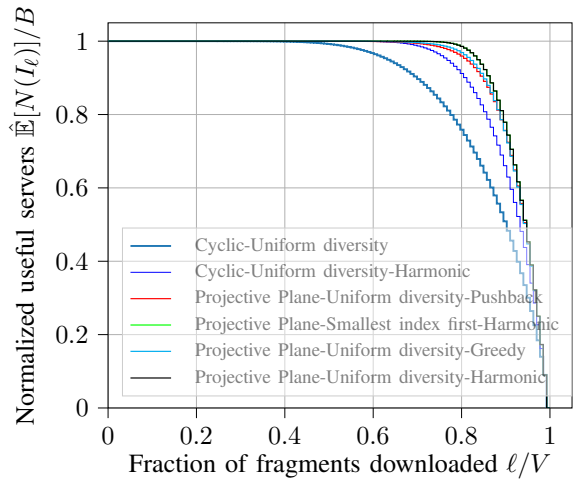}}}
\caption{
This plot shows an empirical average of the normalized number of useful servers $\hat{\mathbb{E}}[N(I_\ell)]/B$ 
for $\frac{12}{133}$-$(133,12)$ replication storage schemes 
based on projective plane and cyclic shift, 
with best nonadaptive scheduling and rank based adaptive scheduling policies.
}
\label{fig:AvgUsefulServerslistForDynamicSchemesExp}
\end{figure}

\begin{figure}[tbh]
\centerline{\scalebox{0.9}{\input{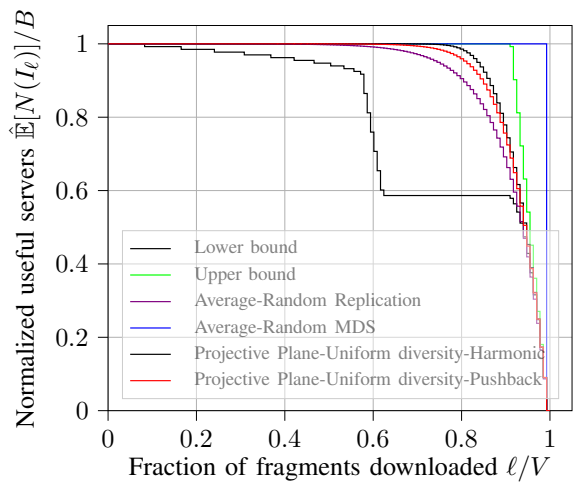}}}
\caption{
Comparison of the best nonadaptive and adaptive scheduling policies for the projective plane based $\frac{12}{133}$-$(133,12)$ replication storage scheme with the normalized universal lower bound in Eq.~\eqref{eq:LB-DesignBased}, 
the normalized upper bound in Eq.~\eqref{eq:UB}, 
and the average performance of $(133,133, 12)$ random replication (Eq.~\eqref{eqn:NumUsefulServersRandom}) and 
random MDS scheme (Eq.~\eqref{eq:N_mds}).
}
\label{fig:ComparisonWithBounds}
\end{figure}

For the following numerical studies, we consider ranked schedulers with greedy and harmonic ranking functions, 
as adaptive scheduling policies. 
In Fig.~\ref{fig:AvgUsefulServerslistForDynamicSchemesExp} these adaptive scheduling policies are compared with the best nonadaptive policies 
for both cyclic shift based and projective plane based storage schemes. 
In this case, we have the freedom to choose the initial schedule, 
since the rank of all fragments remain identical before the first download. 
We explored two types of initial schedules: i) the smallest index first, and  ii) uniform diversity where all fragments are present. 
The performance is very similar in both cases as can be seen in Fig.~\ref{fig:AvgUsefulServerslistForDynamicSchemesExp}, 
with the uniform initialization performing slightly better with respect to the total download time. 
This suggests that for a given storage scheme, the initialization does not affect the overall performance as much.
Note also, that the cyclic storage scheme with adaptive scheduling still does not perform as well as nonadaptive scheduling with design based storage scheme. 
Therefore, it is important to find a good storage scheme. 
As mentioned in Section~\ref{subsec:Cyclic}, the performance of cyclic shift based replication storage with nonadpative scheduling is poor due to large value of overlap parameters.
However, this storage scheme has low overlap for certain fragment and occupancy sets.
An adaptive scheduling exploits this property by driving the system state to ensure the good fragment sets such that the remaining fragments have low overlap in their occupancy sets with low overlap remains.
For cyclic storage with nonadaptive scheduling, spread of overlap parameters should be highlighted, which is exploited by dynamic scheduling over static scheduling. That is, system state is driven towards the good fragment sets, the ones with low overlap.

In Fig.~\ref{fig:ComparisonWithBounds}, we compare the performance of projective plane based storage scheme for 
best performing nonadaptive and adaptive scheduling policies against the bounds, which includes the universal lower bounds on $N(I_\ell)$ derived in Theorems~\ref{thm:LowerBoundNumUsefulLowerGen}~and~\ref{thm:LowerBoundNumUsefulUpperGen}, 
and 
the lower bound on the $N(I_{\ell})$ given in  Eq.~\eqref{eq:LB-DesignBased} for projective plane based storage scheme, the upper bound for $N(I_\ell)$ random replication storage given in Eq.~\eqref{eq:UB}, 
the average number of useful servers of random replication storage ensemble given in Eq.~\eqref{eqn:NumUsefulServersRandom}, 
and the average number of useful servers of random MDS storage ensemble given in Eq.~\eqref{eq:N_mds}~(see Appendix~\ref{app:RandomMDS}).
One important observation is that the performance of the deterministic storage scheme is superior to the average performance of the random code ensemble. 
Therefore, it is worthwhile to develop good deterministic storage schemes. 

\begin{figure}[tbh]
\centerline{\scalebox{1}{\input{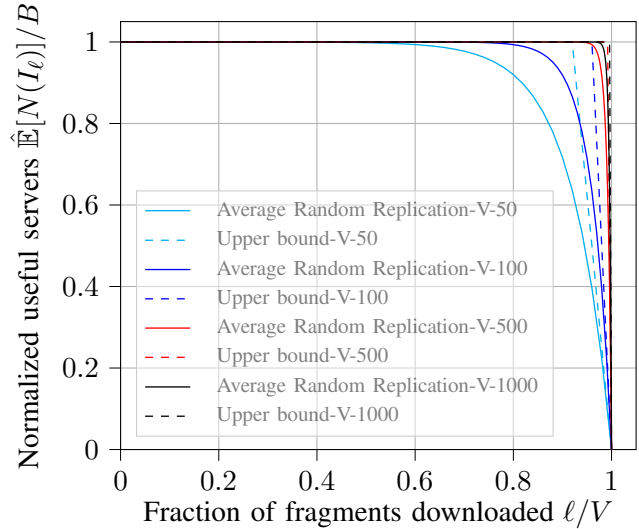}}}
\caption{
We plot the upper bound on the normalized number of useful servers in Eq.~\eqref{eq:UB} and the empirical average of the normalized number of useful servers given in Eq.~\eqref{eqn:NumUsefulServersRandom} for random replication scheme, 
as the number of fragments $V$ increases in the set $\set{50,100,500,1000}$ for a fixed storage capacity per server $\alpha = {K/V} = 0.25$. 
}
\label{fig:AsymptoticAnalytical}
\end{figure}

Finally, in Fig.~\ref{fig:AsymptoticAnalytical}, 
we show the variation of the normalized number of useful servers with increase in the number of servers. 
As indicated in Corollary~\ref{cor:RandomRepAsympOptimal}, 
we observe that the ensemble mean of normalized number of useful servers for the proposed random $(B,V,R)$ replication storage scheme meets the upper bound for any $\alpha$-$(V,R)$ replication storage scheme, 
asymptotically in $V$.

Explicit expression for computation of the mean download time for replication storage schemes is 
given in Eq.~\eqref{eq:MeanDownloadTime}, 
in terms of $\E{1/N(I_\ell)}$ for download $\ell\in\set{0, ,\dots, V-1}$. 
Even for our simpler setup analytical computation of mean download remains elusive. 
We had provided a lower bound on the mean download time in terms of the analytically tractable means $\E{N(I_\ell)}$ for $\ell \in \set{0, \dots, V-1}$. 
We computed the empirical mean of the file download time 
for proposed replication storage schemes and scheduling policies, 
and the corresponding lower bound. 
We observe that for all values of selected system parameters, they remain very close. 
The following Table~\ref{tab:AvgDowloadTimes} shows the empirical mean download times for various storage schemes and scheduling policies.  
\begin{table}[H]
    \centering
    \begin{tabular}{l|l}
		    \hline
            \hspace{1.9cm}\textbf{Storage code} & \hspace{0.25cm} \textbf{Average}  \\
            &\textbf{download time} \\ \hline
            \multicolumn{2}{c}{{Cyclic shift based storage-nonadaptive scheduling}}\\
            \hline
            Uniform diversity & 139629.39 \\
            Smallest index first & 141507.86 \\
            Pushback & 145146.52 \\ \hline
            \multicolumn{2}{c}{{Projective plane based storage-nonadaptive scheduling}}\\
            \hline
            Smallest index first & 122378.76 \\
            Smallest index first-Pushback & 122394.19 \\
            Uniform diversity & 122897.40 \\
            Uniform diversity-Pushback & 121678.81 \\ \hline
            \multicolumn{2}{c}{{Projective plane based storage-adaptive scheduling}}\\
            \hline
			Smallest index first-Harmonic ranking & 121002.69 \\
			Uniform diversity-Harmonic ranking & 120886.04 \\
            Smallest index first-Pushback-Harmonic ranking & 120940.41 \\
            Uniform diversity-Pushback-Harmonic ranking & 120993.85 \\
            Uniform diversity-Greedy & 121617.66  \\ \hline
            \multicolumn{2}{c}{{Cyclic shift based storage-adaptive scheduling}}\\
            \hline
			Uniform diversity-Harmonic ranking & 126722.19 \\
			Smallest index first-Harmonic ranking & 126769.84 \\
			Pushback-Harmonic ranking & 126783.50 \\ \hline
		\end{tabular}
		\caption{Average download times of various nonadaptive and adaptive policies.}
    \label{tab:AvgDowloadTimes}
\end{table}

{\em Discussion.}  
Storage schemes based on combinatorial designs perform better than the naive schemes such as the cyclic shift based storage schemes. 
The scheduling policy significantly affects the performance for the same storage code. 

For any storage code, 
adaptive scheduling leads to better performance over nonadaptive scheduling. 
The initial set of fragments placed at the head of the servers does not noticeably affect the performance of the adaptive scheduling algorithms. 
Adaptive scheduling also reduces the dependence on the storage code as can be seen that both cyclic and design based storage schemes lead to comparable performance. 

\section{Conclusion}
\label{sec:Conclusion}

\subsection{Summary}
We considered a single file that is divided into finitely many fragments. 
The fragments are identically replicated and the replicas are stored on a finite server system with finite storage capacity. 
A file is considered downloaded, if one can reconstruct the file from the downloaded fragments. 
We posed the problem of optimal storage scheme and scheduling policy to minimize the mean download time of the entire file, 
given that each fragment download time is random \iid and memoryless. 
We modified the problem from the minimization of mean download time to the maximization of the aggregate number of useful servers over all fragment downloads. 
We subdivided this problem into two subproblems. 
The first subproblem was to find the optimal scheduling policy given a storage scheme. 
The second subproblem was to find the optimal storage scheme for a given scheduling policy. 
We provided lower bounds on the number of useful servers, 
and proposed design based storage schemes that maximize this lower bound. 
Further, we posed the optimal adaptive scheduling policy as an MDP, and provided suboptimal solutions. 
We empirically verified that the proposed solution is close to the upper bound on the number of useful servers, 
when the number of fragments is large. 

\subsection{Discussion and further directions}
Our storage scheme and scheduling policy can be generalized to other codes as well. 
The problem of optimal storage and scheduling policy exists for all coded storage schemes, except for MDS coded storage. For MDS coded storage scheme, all fragments are useful until $V$ coded fragments are downloaded.  
We showed that MDS coded storage minimizes the mean download time. 
However, we observed that either when the servers have sufficiently large storage or when the number of fragments is large, 
the mean download time performance of replication coded storage is competitive to that of MDS coded storage. 

We note that the exponential service distribution assumption was needed to compute the expression for mean download time in terms of number of useful servers. 
Our study can be extended to any service distribution that leads to a smaller mean download time for a more dominant sequence of number of useful servers. 
{
The mean download time for a general distribution for fragment download time can also be minimized if it was a decreasing function of the sequence of number of useful servers. 
}
In this case, the design based storage schemes will continue to do well, since they attempt to maximize the number of useful servers. 
However, for non-memoryless distribution, the scheduling policies are more complicated, 
since one needs to take care of age of previously scheduled fragment replicas, which are not yet downloaded. 

We also note that, even though our study was for a single file, our proposed framework can be extended to multiple files. 
Specially, when we assume that each server has a predetermined fraction of storage for each file, 
one can find optimal storage scheme for each of these files separately. 
However, the scheduling for multiple files become somewhat more involved in this case. 
One has to schedule a fragment at each server that could be useful for one of these files. 
For a specific file, which fragment to schedule can be selected by one of our proposed scheduling algorithms. 
However, it is not clear \emph{apriori}, 
which one of the files should be scheduled. 

In our studies, we have ignored the delays in cancelling the scheduled replicas. 
We observe that these delays affect other coding policies as well. 
There is a subtle difference though. 
Each download for a replication coded storage, leads to cancellation at all other servers where the identical replica was being downloaded. 
This can lead to a maximum of $R$ cancellations per download. In MDS coded storage, there are no cancellations until the $V$th download, when every other scheduled fragment at other $N(I_{V-1})-1$ servers should be cancelled. 

As can be seen from the preceding discussion, there are many interesting directions of practical import to explore; 
we hope that this motivates further research in this topic. 

	\begin{appendices}

\section{Proof of Lemma~\ref{prop:design prelims}}
\label{app:designs}

For the first conservation law, we observe that 
\EQ{
VR =\sum_{v \in \cP}\sum_{S \in \cB}\SetIn{v \in S} 
= \sum_{S \in \cB}\sum_{v \in \cP}\SetIn{v \in S} 
= BK.
}
For the second conservation law, we observe that 
\EQ{
\lambda\binom{V}{t} 
= \sum_{W \in \binom{\cP}{t}}\sum_{S \in \cB}\SetIn{W \subseteq S}
}
where, $\binom{V}{t}$ denotes the number of $t$-subsets that can be chosen from the set $[V]$. 
Changing the order of the summation, 
we can write the RHS of the above equation as 
\EQ{
\sum_{S \in \cB}\sum_{W \in \binom{\cP}{t}}\SetIn{W \subseteq S} 
= \sum_{S \in \cB}\sum_{W \in \binom{[S]}{t}}1 = B\binom{K}{t}.
}

\section{MDP results}

\subsection{Proof of Lemma~\ref{lem:SchedulingCTMC}}
\label{app:SchedulingCTMCProof}

\begin{IEEEproof}
A countable state process with sample paths that are right continuous with left limits, 
is Markov if 
(a) the inter-transition times are memoryless, 
(b) conditioned on the current state the future inter-transition times are independent of the past, and 
(c) the jump probabilities depend only on the current state~\cite[Chapter 5]{Ross1995Wiley}. 
We will show that the countable state process $X$ satisfies all three conditions given a completely utilizing replication storage scheme $\Phi \in \cS$ and associated work conserving scheduling policy $\Psi_\Phi$.  

Recall that each sample path of the process $X$ is piece-wise constant, and transitions only at the download instants. 
Thus, the process $X$ is right continuous with left limits. 
After the $\ell$th download, 
the time to download the next fragment $v_{\ell + 1}$ is the minimum of the residual download time at each of the $N(I_\ell)$ useful servers given the current state $I_{\ell}$.
As the service times at all servers are  \iid and exponentially distributed with rate $\mu$, it follows that the time for next download $D_{\ell+1} - D_{\ell}$ is exponentially distributed with rate $N(I_\ell)\mu$. 
The memoryless property of the service times also implies that the residual download time at each useful server is independent of the past. 
Thus, future inter-transition times $(D_{j+1}-D_j: j \ge \ell)$ are independent of the past conditioned on the current state $I_\ell$.  
In addition, since residual fragment download times are identically exponentially distributed, 
it follows that probability of any of the useful servers finishing first is $\frac{1}{N(I_\ell)}$. 
Since the scheduled fragment on each useful server depends only on the current state $I_\ell$, 
the transition probability from current state $I_{\ell}$ to the next state $I_{\ell+1}$ 
depends only on the current state. 
This transition probability is denoted
\EQ{
p_{ I_{\ell}, I_{\ell}\cup\set{v}} = \frac{\sum_{b \in U(I_\ell)}\SetIn{\Psi_\Phi(I_\ell)(b)=v}}{N(I_\ell)}. 
}
We observe that the three conditions outlined above are met by the process $X$ and hence, the result holds.
\end{IEEEproof}

\subsection{Backward induction algorithm}
\label{app:BackwardInduction}
For completeness, we discuss the backward induction algorithm which will provide us the optimal work conserving scheduling policy $\Psi_{\Phi}^\ast$ for a fixed completely utilizing $\alpha$-$(V,R)$ replication storage policy $\Phi$.
Recall that the utility function for the decision rule $\Psi_\Phi(I_{\ell-1})$ is the reward-to-go from the remaining downloads, and is given by
\eq{
u_{\ell}(\Psi_{\Phi}(I_{\ell-1}))  
=& \sum_{v \notin I_{\ell-1}} p_{ I_{\ell-1}, I_{\ell-1}\cup\set{v}} \Big[r_{\ell}(I_{\ell-1}\cup\set{v}) + u_{\ell+1}(\Psi_{\Phi}(I_{\ell-1}\cup\set{v}) \Big].
}
A file is considered downloaded when all $V$ fragments have been downloaded, and hence we restrict our study to transition instants until $D_{V-1}$. 
Accordingly, we set the terminal utility function $u_V(\Psi_{\Phi}(I_{V-1}))$ to be zero.
Then, the backward induction can be performed as follows. 

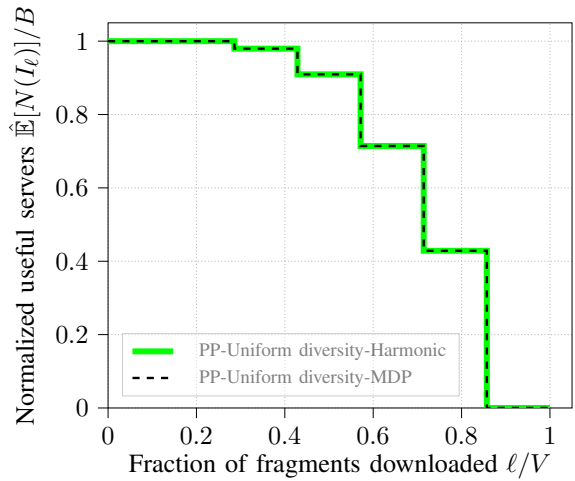
\begin{figure}[hhh]
\centerline{\scalebox{0.9}{% This file was created by tikzplotlib v0.9.1.
\begin{tikzpicture}

\definecolor{color0}{rgb}{0.12156862745098,0.466666666666667,0.705882352941177}
\definecolor{color1}{rgb}{1,0.498039215686275,0.0549019607843137}
\definecolor{color2}{rgb}{0.172549019607843,0.627450980392157,0.172549019607843}
\definecolor{color3}{rgb}{0.83921568627451,0.152941176470588,0.156862745098039}
\definecolor{color4}{rgb}{0.580392156862745,0.403921568627451,0.741176470588235}
\definecolor{color5}{rgb}{0.549019607843137,0.337254901960784,0.294117647058824}

\begin{axis}[
legend cell align={left},
legend style={fill opacity=0.5, draw opacity=1, text opacity=1, at={(0.03,0.03)}, anchor=south west, draw=white!80.0!black, font=\scriptsize},
tick align=outside,
tick pos=left,
x grid style={white!69.01960784313725!black, densely dotted},
xlabel={Fraction of fragments downloaded $\ell/V$},
xmajorgrids,
xmin=0, xmax=1.05,
xminorgrids,
xtick style={color=black},
y grid style={white!69.01960784313725!black, densely dotted},
ylabel={Normalized useful servers $\hat{\mathbb{E}}[N(I_\ell)]/B$},
ymajorgrids,
ymin=0, ymax=1.05,
yminorgrids,
ytick style={color=black}
]
% ---- q=2 Start----
\addplot [green, const plot mark right, mark=, mark size=1, mark options={solid}, line width=2.5pt]
table {%
0	1
0.142857143	1
0.285714286	1
0.428571429	0.979618571
0.571428571	0.909615
0.714285714	0.714285714
0.857142857	0.428571429
1	0
};
\addlegendentry{PP-Uniform diversity-Harmonic}

\addplot [black, dashed, const plot mark right, mark=, mark size=1, mark options={solid}, line width=1pt]
table {%
0	1
0.142857143	1
0.285714286	1
0.428571429	0.979615429
0.571428571	0.909697429
0.714285714	0.714285714
0.857142857	0.428571429
1	0
};
\addlegendentry{PP-Uniform diversity-MDP}
% ---- q=2 End----

% ---- q=3 Start----
%\addplot [green, const plot mark right, mark=, mark size=1, mark options={solid}, line width=1pt]
%table {%
%0	1
%0.076923077	1
%0.153846154	1
%0.230769231	1
%0.307692308	0.999502308
%0.384615385	0.997786923
%0.461538462	0.991493077
%0.538461538	0.973373846
%0.615384615	0.933525385
%0.692307692	0.850932308
%0.769230769	0.721257692
%0.846153846	0.538461538
%0.923076923	0.307692308
%1	0
%};
%\addlegendentry{q=3-PP-Uniform diversity-Harmonic}
%
%\addplot [black, dashed, const plot mark right, mark=, mark size=1, mark options={solid}, line width=0.5pt]
%table {%
%0	1
%0.076923077	1
%0.153846154	1
%0.230769231	1
%0.307692308	0.999502308
%0.384615385	0.997786923
%0.461538462	0.991493077
%0.538461538	0.973373846
%0.615384615	0.933525385
%0.692307692	0.850932308
%0.769230769	0.721257692
%0.846153846	0.538461538
%0.923076923	0.307692308
%1	0
%};
%\addlegendentry{\red{q=3-PP-Uniform diversity-MDP}}
% ---- q=3 End----

\end{axis}

\end{tikzpicture}}}
\caption{
%\red{[q=3-PP-Uniform diversity-MDP plot is pending!]}
This plot shows an empirical average of the normalized number of useful servers $\hat{\mathbb{E}}[N(I_\ell)]/B$ for the projective plane (PP) based $\frac{3}{7}$-$(7,3)$, %\red{$\frac{4}{13}$-$(13,4)$} 
replication storage scheme with optimal backward induction algorithm.
}
\label{fig:MDP}
\end{figure}

\begin{enumerate}
\item 
At the terminal decision epoch $D_{V-1}$, 
the only remaining fragment to be downloaded is $\set{v} = [V] \setminus I_{V-1}$. 
This fragment is stored on the set of servers $\Phi_v$, 
which are the only useful servers with remaining fragment sets $(S_b^{V-1}= \set{v}: b \in \Phi_v)$. 
Therefore, the scheduling decision on these servers is trivially given $\Psi_\Phi(I_{V-1})(b) = v$ for all $b \in \Phi_v$, 
and $u^\ast_V(I_{V-1}) = 0$. 
\item 
At the download instant $D_{V-2}$, we have
\EQ{
u_{V-1}(\Psi_{\Phi}(I_{V-2})) 
= \sum_{v \notin I_{V-2}} p_{I_{V-2}, I_{V-2}\cup \set{v}}\frac{N(I_{V-1})}{V}. 
} 
As $N(I_{V-1}) = \abs{\Phi_{v_V}} =  R$, we get
\EQ{
u_{V-1}^\ast(I_{V-2}) = \frac{R}{V}. 
}
It follows that the utility function is independent of the chosen scheduling policy, 
and any arbitrary scheduling decision can be made at this stage. 
\item Given the scheduling decisions at download instants $\set{D_{\ell+1}, \dots, D_{V-1}}$, we consider the utility function at download instant $D_\ell$.  
We calculate
\eq{
u_{\ell+1}^\ast(I_\ell) 
=& \max_{\Psi_{\Phi}} \Bigg[\sum_{v \notin I_{\ell}} p_{ I_{\ell}, I_{\ell}\cup\set{v}} \Big[r_{\ell+1}(I_{\ell}\cup\set{v}) +u_{\ell+2}^*(I_{\ell} \cup\set{v})  \Big] \Bigg].
}
Also, we get the optimal scheduling decision at the current epoch to be
\eq{
\Psi_\Phi^\ast(I_{\ell}) =& \arg \max_{\Psi_{\Phi}} \Bigg[\sum_{v \notin I_{\ell}} p_{ I_{\ell}, I_{\ell}\cup\set{v}} \Big[r_{\ell+1}(I_{\ell}\cup\set{v}) +u_{\ell+2}^*(I_{\ell} \cup\set{v})  \Big] \Bigg].
}
For a given download subsequence $I_{\ell}$, the scheduling decision at each useful server $b \in U(I_\ell)$ is $\Psi_{\Phi}(I_{\ell})(b)$. 
This decision affects the transition probabilities $p_{ I_{\ell}, I_{\ell}\cup\set{v}}$ for all remaining fragments $v \in I_{\ell}^c$.
Since the optimal reward-to-go function $u_{\ell+2}^*(I_{\ell} \cup\set{v})$ from the next decision epoch $D_{\ell+1}$  is already available from the previous step,
the optimal scheduling decision at each server can be chosen to schedule the fragment that maximizes the reward-to-go function $u_{\ell+1}^\ast(I_\ell)$. 

\item If the present decision epoch is $ D_{0}$, stop. Else, return to previous step.
\end{enumerate}

\subsection{Computation complexity of backward induction algorithm}
\label{subsec:CompComplexMDP}
At every inductive step, the backward induction algorithm requires us to compute and record the optimal scheduling decisions for every possible download subsequence $I_{\ell}$ and the corresponding reward-to-go function.
At every $\ell$th download instant the number of possible download subsequences are $\binom{V}{\ell}$ and hence the total number of evaluations of the reward-to-go function summed up over the total number of stages grows exponentially with the number of fragments.

We observe that this approach is computationally very expensive in comparison to the ranked schedulers.
However, in Fig.~\ref{fig:MDP}, we provide the backward induction algorithm results for small number of fragments $V=7$ of the projective plane (PP) based $\frac{3}{7}$-$(7,3)$ scheme with initial arrangement as in Table~\ref{tab:design-matching} and ranked scheduler given in  Algorithm~\ref{Algo:SuboptimalScheduler} with harmonic ranking function.
Thus, we demonstrate that ranked schedulers with suitable rank function can provide near optimal performance in certain cases of interest.

\section{Randomized placement of MDS codes} 
\label{app:RandomMDS}
In this section, we study a randomized $(B,V,R)$ MDS code analogous to the randomized $(B,V,R)$ replication storage scheme defined in Definition~\ref{defn:RandomPlaceRep}.
\begin{defn}
\label{defn:RandomPlaceMDS}
A randomized $(B,V,R)$ MDS code is defined as a $(VR,V)$ MDS code stored on a system with $B$ servers each having a storage capacity of $VR$ fragments. 
Each coded fragment $v \in [VR]$ is stored on a server $\chi_v \in[B]$ independently with uniform probability $1/B$.
The collection of all random $(B,V,R)$ MDS replication storage schemes is referred to as the random $(B,V,R)$ MDS storage ensemble.
\end{defn}
\nix{\red{Hence, to gain some insight }
we consider the following random placement of $(VR, V)$ MDS code on $B$ servers which is asymptotically equivalent to $(B,V,R,K)$ placement. 

\begin{defn}
We define a $(VR, V)$ MDS storage policy where each coded fragment $v \in [VR]$ is stored on a server $\chi_v \in[B]$ independently with uniform probability $1/B$. 
\end{defn}
}

We denote the normalized number of fragments stored on a server $b \in [B]$ as $\alpha_b^{\mds} \triangleq \frac{1}{V}\sum_{v \in [VR]}\SetIn{\chi_v = b}$. 
Further, as for the randomized replication storage scheme, we say that a randomized $(B,V,R)$ MDS storage scheme is an $\alpha$-$(V,R)$ storage scheme asymptotically in $V$, if for each server $b$, $\lim_{V \to \infty} \alpha_b^{mds} = \alpha$ almost surely.
\begin{thm}
\label{thm:mds_serverstorage}
The randomized $(B,V,R)$ storage scheme defined in Definition~\ref{defn:RandomPlaceMDS} is an $\alpha$-$(V,R)$ storage scheme asymptotically in $V$. 
\end{thm}
\begin{IEEEproof}
The proof is very similar to that of Theorem \ref{thm:replication_serverstorage}. 
We consider the sequence $(M_n: n \le V)$ where
\EQ{
M_n \triangleq \frac{1}{V}\sum_{v =1}^{n} (\SetIn{\chi_{v} = b} - \E{\SetIn{\chi_{v} = b}}),~n \le VR.
}
Since the summands are \iid and zero mean, the sequence $(M_n: n \le V)$ can be verified to be a Martingale.
In addition,
\EQ{
\E{M_n^2} = \frac{n}{V^2} \Var(\SetIn{\chi_v} = b) = \frac{n}{BV^2}(1 - \frac{1}{B}).
}
Then, we can bound the limiting mean of absolute value of martingale $M_n$ as in the proof of Theorem \ref{thm:replication_serverstorage} as 
\eq{
0 \le \limsup_{V \to \infty}\sup_{n \le VR}\E{\abs{M_n}} &\le \limsup_{V \to \infty} \sup_{n \le VR}\sqrt{\E{M_n^2}} 
\le 0.
}
Thus,
we conclude that $M_V$ converges to $0$ almost surely and result follows from the observation that $M_{VR} = \alpha_b^{mds} - \alpha$. 
\end{IEEEproof}
\begin{thm}
\label{thm:performance_random_mds}
For the random $(B,V,R)$ MDS storage ensemble defined in Definition~\ref{defn:RandomPlaceMDS}, we have
\EQ{
\frac{1}{BV}\sum_{\ell = 0}^{V-1} \E{N(I_{\ell})} = 1 - \frac{B}{V}(1 - \frac{1}{B})^{(R-1)V+1} \Big(1- (1 - \frac{1}{B})^V\Big).
}
\end{thm}
\begin{IEEEproof} 
The proof is analogous to the proof of Theorem~\ref{thm:performance_random_rep}.
The set and number of useful servers after $\ell$ downloads are
$U(I_{\ell}) = {\cup_{v \notin I_\ell} \chi_v}$ and $N(I_\ell) = \abs{U(I_\ell)}$ respectively. 

The probability of event $\set{\chi_v \neq b}$ is given by $(1- \frac{1}{B})$ for any fragment $v \in [VR]$ and any server $b \in [B]$. 
Since each placement was independent and uniform, we get 
\EQ{
P\set{b \in U(I_\ell)} = 1 - \Big(1 - \frac{1}{B}\Big)^{RV - \ell}.
}
Since $N(I_\ell) = \sum_{b \in [B]}\SetIn{b \in U(I_\ell)}$, 
it follows from the linearity of expectation that 
\EQN{
\label{eq:N_mds}
\E{N(I_{\ell})} = B\Big[1 - \Big(1 - \frac{1}{B}\Big)^{RV - \ell}\Big].
}
Result follows from summing the above equation on both sides for $\ell \in \set{0, \dots, V-1}$ and normalizing by $BV$. 
\end{IEEEproof}

\begin{rem}
If $x \triangleq (1 - \frac{1}{B})^{-1} \ge 1$, then from Theorem~\ref{thm:performance_random_rep}, we get
\eqn{
\label{eq:RandomRepApprox}
\frac{1}{V}\ln(1 - \frac{1}{BV}\sum_{\ell=0}^{V-1}\E{N(I_\ell)}) 
&= - R\ln x + \frac{1}{V}\ln\frac{x^{VR}-1}{x^R-1}\approx \alpha + \frac{1}{V}\ln VR
}
for the random $(B,V,R)$ replication storage ensemble.
Similarly, from Theorem~\ref{thm:performance_random_mds}, we obtain that for a random $(B,V,R)$ MDS storage ensemble  
\eqn{
\label{eq:RandomMDSApprox}
\frac{1}{V}\ln(1 - \frac{1}{BV}\sum_{\ell=0}^{V-1}\E{N(I_\ell)}) 
&= - R \ln x +  \frac{1}{V}\ln\frac{x^V-1}{x-1}\approx \alpha + \frac{1}{V}\ln V.
}
We infer from Eq.~\eqref{eq:RandomRepApprox} and Eq.~\eqref{eq:RandomMDSApprox} that the performance of the random $(B,V,R)$ replication storage schemes is comparable to that of random $(B,V,R)$ MDS storage schemes in the limit of large number fragments.
\end{rem}

\end{appendices}
	
	\bibliographystyle{IEEEtran}
	\bibliography{IEEEabrv,tit-2020}
\end{document}